\pdfoutput=1
\documentclass[sigconf]{acmart}

\AtBeginDocument{%
  \providecommand\BibTeX{{%
    \normalfont B\kern-0.5em{\scshape i\kern-0.25em b}\kern-0.8em\TeX}}}

\setcopyright{acmcopyright}
\copyrightyear{2022}
\acmYear{2022}
\acmDOI{}

\newcount\COMMENTs  
\usepackage{color}
\usepackage{algorithm}
\usepackage{algpseudocode}
\usepackage{enumitem}
\definecolor{darkgreen}{rgb}{0,0.5,0}
\definecolor{purple}{rgb}{1,0,1}
\newcommand{\comm}[2]{\ifnum\COMMENTs=1\textcolor{#1}{#2}\fi}
\newcommand{\sun}[1]{\textcolor{purple}{#1}}

\newcommand{\hide}[1]{}

\setcopyright{acmcopyright}
\copyrightyear{2023}
\acmYear{2023}
\acmDOI{XXXXXXX.XXXXXXX}

\acmConference[Conference acronym 'XX]{Make sure to enter the correct
  conference title from your rights confirmation email}{June 03--05,
  2023}{Woodstock, NY}
\acmPrice{15.00}
\acmISBN{978-1-4503-XXXX-X/23/06}

\usepackage{makecell}
\usepackage{multicol}
\usepackage{subcaption}
\usepackage{siunitx}

\usepackage{amsmath,amsfonts,bm}

\def\eqref#1{equation~\ref{#1}}

\def\1{\bm{1}}

\def\rvx{{\mathbf{x}}}

\def\vmu{{\bm{\mu}}}

\def\mSigma{{\bm{\Sigma}}}

\DeclareMathAlphabet{\mathsfit}{\encodingdefault}{\sfdefault}{m}{sl}
\SetMathAlphabet{\mathsfit}{bold}{\encodingdefault}{\sfdefault}{bx}{n}

\def\gR{{\mathcal{R}}}

\begin{document}

\newcommand{\model}{GraphVF}

\title{\model: Controllable Protein-Specific 3D Molecule Generation with Variational Flow}

\author{Fang Sun}
\email{fts@pku.edu.cn}
\affiliation{
  \institution{Peking University}
  \city{Beijing}
  \country{China}
}

\author{Zhihao Zhan}
\email{zhan8855@pku.edu.cn}
\affiliation{
  \institution{Peking University}
  \city{Beijing}
  \country{China}
}

\author{Hongyu Guo}
\email{hongyu.guo@uottawa.ca}
\affiliation{
  \institution{National Research Council Canada}
  \institution{University of Ottawa}
  \city{Ottawa}
  \country{Canada}
}

\author{Ming Zhang}
\email{mzhang\_cs@pku.edu.cn}
\affiliation{
  \institution{Peking University}
  \city{Beijing}
  \country{China}
}

\author{Jian Tang}
\email{jian.tang@hec.ca}
\affiliation{
  \institution{Mila - Qu\'ebec AI Institute}
  \institution{HEC Montr\'eal}
  \institution{CIFAR AI Research Chair}
  \city{Montr\'eal}
  \country{Canada}
}

\renewcommand{\shortauthors}{Sun, et al.}

\begin{abstract}
Designing molecules that bind to specific target proteins is a fundamental task 
in drug discovery. 
Recent models leverage geometric constraints to generate ligand 
molecules that bind cohesively with specific protein pockets. However, these 
models 
cannot effectively generate 3D molecules with 2D skeletal curtailments 
and property constraints, which are pivotal to 
drug potency and development. 
To tackle this challenge, we propose \model, a variational flow-based framework that combines 2D topology and 3D geometry, for controllable generation of binding 3D molecules. 
Empirically, our method achieves state-of-the-art binding affinity and 
realistic sub-structural layouts for protein-specific generation. 
In particular, \model\ represents the first \textit{controllable} 
geometry-aware, protein-specific molecule generation method, which 
can generate binding 3D molecules with tailored 
sub-structures and physio-chemical properties. 
Our code is available at \href{https://github.com/Franco-Solis/GraphVF-code}{https://github.com/Franco-Solis/GraphVF-code}. 

\end{abstract}

   
 

\maketitle

\section{Introduction}

The \textit{de novo} design of synthetically feasible drug-like molecules that 
bind to specific protein pockets is a crucial yet very challenging task in drug 
discovery. To cope with such challenges, there has been a  recent surge of interest 
in leveraging deep generative models to effectively search the chemical space for 
molecules with desired properties. These machine-learning models typically encode 
the chemical structures of molecules into a low-dimensional space, which then can 
be optimized and sampled to generate potential 2D or 3D molecule candidates~\citep{
jin2018junction,shi2020graphaf,zhu2022direct,hoogeboom2022equivariant}. Along this 
research line, a more promising direction has also been explored recently: generating 
3D molecules that bind to given protein pockets. 

Such protein-specific 3D molecule generation is fundamentally important because it 
touches down to the micro-scale principle of drug functionalities: how the 
drug ligand and the receptor protein interact with each other in the 3D space. 
Current state-of-the-art generative models~\citep{luo20213d,peng2022pocket2mol, liu2022generating} leverage   equivariant 3D graph neural networks 
to 
generate drug molecules (i.e., ligands) directly based on the 3D geometry of 
the binding pocket. They explicitly capture the fine-grained 3D atomic 
interactions and produce ligand poses that directly fit into the given binding pocket. Nevertheless, two critical issues 
remain unsolved for these existing geometric approaches: 

\textit{1) Enforcing 2D Pharmacophoric Pattern in 3D generation. } 
2D molecular scaffolds embody rich 
pharmacophoric patterns, 
determining ligands' bio-chemical activities 
and binding affinity to a large extent~\citep{WermuthGanellinLindbergMitscher+1998+1129+1143}. 
Consider, for example, the molecules of serotonin (a benign neurotransmitter) and 
N,N-Dimethyltryptamine (DMT, a famous hallucinogen). As shown in Figure
~\ref{figure:pharmacophore}a of Appendix~\ref{app:chem-struc-illus}, serotonin 
and DMT share a large common bulk of their structures, which both possess an indole 
and an ethylamine group, but differ enormously in their neural activities. In 
fact, the extra Methyl groups in DMT's $\mathrm{NHMe_2}$ are pharmacophoric, 
inducing an attractive charge interaction with Asp-231~\citep{gomez2015note}. 
This pharmacophoric feature gives rise to DMT's binding affinity with the 
5-$\mathrm{HT_{2A}}$ binding site and produces hallucination (Figure
~\ref{figure:pharmacophore}b). Such observations 
suggest that effectively enforcing 2D pharmacophoric patterns in 3D ligands is critical 
for binding. 
Nevertheless, existing 3D molecular generation methods do not take into account these useful 2D patterns. 

\textit{2) Controllable generation. } When it comes to drug manufacturing, 
physio-chemical properties like solubility, polarizability, and toxicity are 
equally important to drug quality. We have to effectively control these qualities 
to make sure that the synthesized drug molecules have good exposure, \textit{e.g.} absorption, distribution, metabolism, excretion (ADME) \textit{in vivo}, and thus sufficient efficacy in clinical trials~\citep{egan2010predicting}. 
It is worth noting that, although recent models like EDM~\citep{hoogeboom2022equivariant} 
and RetMol~\citep{wang2022retrieval}
have been popular for their capability to perform controlled generation on these properties, performing 
controllable generation in the 3D pocket-specific scenario remains unexplored by previous works. 

\begin{figure}[!ht]
   \centering
   \includegraphics[width=0.95\linewidth]{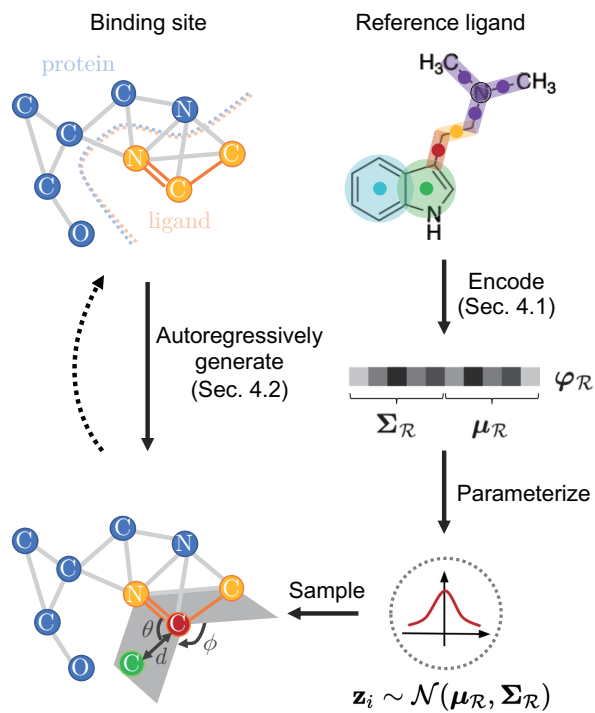}
   \caption{Overview of the \model{} variational flow workflow. \model{} autoregressively samples from the variational distribution (bottom right) to generate the atoms and bonds of the drug ligand at the binding site using flow transformation (left column). The flow function is parameterized by the 3D binding site geometry. The variational distribution is encoded from the 2D topology of the reference ligand during training (right column). During controllable generation, the reference ligand is not provided. The latent encoding $\varphi_\mathcal{R}$ is calculated by mean aggregation of molecules with specified qualities from the training set or selected using Bayesian Optimization, which achieves controllability. }
   \label{figure:overview}
\end{figure}

To address the aforementioned two issues, we propose \model {} (Graph Variational Flow), a protein-specific 
molecule generation framework. 
In a nutshell, \model{} leverages a flow-based generative model to integrate both 3D geometric and 2D 
topological 
constraints, aiming to gain control over the structure and property of the generated 3D 
ligands. 
In the generative process, the flow model here  transforms a simple prior distribution into the complex distribution of the 3D binding molecules.  As illustrated in Figure~\ref{figure:overview}, to  obtain desirable topological information in the generated 3D molecules, the \model{} method encodes  the 2D molecular topology  patterns into such prior distribution. 
Consequently, sampling from such prior 
enables the flow model to facilitate the generation of 3D molecules with specific sub-structures and physio-chemical properties encoded in the prior distribution.

We show empirically that \model{} can generate novel 3D drug molecules with high binding 
affinity to the receptor proteins, 
outperforming state-of-the-art methods by a large margin. Also, the generated molecules have 
more realistic sub-structure distribution, demonstrating the effectiveness of combining 
3D geometry and 2D topology in protein-specific generation tasks. 
More importantly, \model\ exposes a clean-cut VAE-style interface 
for imposing customized constraints, which is extremely useful in practice for 
controlling the sub-structure and physio-chemical properties of the generated drug ligands. 

Our main contributions are summarized as follows. 
\begin{itemize}
\setlength{\itemsep}{0pt}
\setlength{\parsep}{0pt}
\setlength{\parskip}{0pt}
\item We introduce the first method that enables generating 3D molecules with specified
chemical sub-structures or physio-chemical properties.
\item We devise a novel variational flow-based framework to effectively integrate 3D
geometric and 2D topological constraints for protein-specific 3D molecule generation.

\item We empirically demonstrate our method's superior performance to  state-of-the-art 
approaches in generating 3D binding molecules. 
\end{itemize}

\section{Related Work}
With the development of geometric deep learning and probabilistic generative models in the 
past few years, \textit{de novo} molecule generation techniques have evolved drastically, empowering us to 
sample diverse high-quality molecules with desired properties under various complex scenarios. To this end, there are three major tasks: 2D molecule generation, 3D unbounded molecule generation, and 3D 
protein-specific molecule generation. 


\paragraph{2D molecule generation. } This task~\citep{you2018graph, simonovsky2018graphvae, 
jin2018junction, liu2018constrained, madhawa2019graphnvp, shi2020graphaf} aims at mining 
high-quality topological representations and generating valid 2D molecular graphs from scratch. For example, 
GraphAF~\citep{shi2020graphaf} uses a flow-based model to generate atoms and bonds in an 
autoregressive manner. JT-VAE~\citep{jin2018junction} generates molecular graphs with the 
guidance of a tree-structured scaffold over chemical substructures. To optimize molecules 
toward desired properties, models like GCPN~\citep{you2018graph} and GraphAF~\citep{shi2020graphaf} 
adopt reinforcement learning to tune the model. For variational auto-encoder (VAE) based 
models like SD-VAE~\citep{dai2018syntax} and JT-VAE~\citep{jin2018junction}, each latent encoding 
of the variational distribution corresponds to a specific group of molecules in the chemical space. 
Therefore, these VAE-based models can perform zero-shot optimization without retraining the model, 
as long as they can acquire the optimal latent embedding via linear regression, bayesian optimization, etc. 

\paragraph{3D Non-Protein Specific Molecule Generation} 
This task~\citep{luo2022autoregressive, li2021learning, hoogeboom2022equivariant} aims to learn the 
geometric representations of molecules in the 3D space and generate valid molecules with reasonable 
3D conformation. For instance, G-SphereNet~\citep{luo2022autoregressive} uses symmetry invariant 
representations in a spherical coordinate system (SCS) to generate atoms in the 3D space and preserve 
equivariance. L-Net~\citep{li2021learning} encodes hierarchical molecular structure with Graph U-Net and directly outputs the topology and geometry of the molecule through a valency rule-based backtracking algorithm. EDM~\citep{hoogeboom2022equivariant} is an equivariant diffusion model that 
generates 3D molecule geometry via an iterative denoising process. EDM can be configured to perform 
controllable generation over certain property $c$ by re-training the diffusion model with $c$'s feature vector concatenated to its E(n) equivariant dynamics function. RetMol~\citep{wang2022retrieval} is a retrieval-based framework for controllable molecule generation that requires no task-specific fine-tuning. 


\paragraph{3D Molecule Generation for Target Protein Binding}
With the wide availability of large-scale datasets~\citep{francoeur2020three, li2021structure} 
for target protein binding, recent works~\citep{peng2022pocket2mol, liu2022generating, lin2022diffbp, 
schneuing2022structure} have been able to generate drug ligands 
directly based on the 3D geometry of the binding pockets. For example, Pocket2Mol
~\citep{peng2022pocket2mol} leverages a spatial-autoregressive model; it directly models 
the \textit{p.d.f.} for atom occurrence in the 3D space as a Gaussian mixture (GMM), 
and then iteratively places the atoms from the learned distribution until there 
is no room for new atoms. GraphBP~\citep{liu2022generating}, an autoregressive 
model, retains good model capacity via normalizing flow; variables are randomly 
sampled from a compact latent space, before they are projected into the chemical 
space by an arbitrarily complex flow transformation. DiffBP~\citep{lin2022diffbp} 
considers the global interaction between the protein pocket and the ligand molecule, and uses 
a diffusion model to generate ligand molecules non-autoregressively. Despite the promising potential 
along this line of purely geometric approach, these methods cannot explicitly perceive the topological 
pharmacophoric patterns within the ligand structure. Nor can they conduct explicit control over 
specific chemical sub-structures and physio-chemical properties. 

\begin{figure}[!ht]
   \centering
   \includegraphics[width=\linewidth]{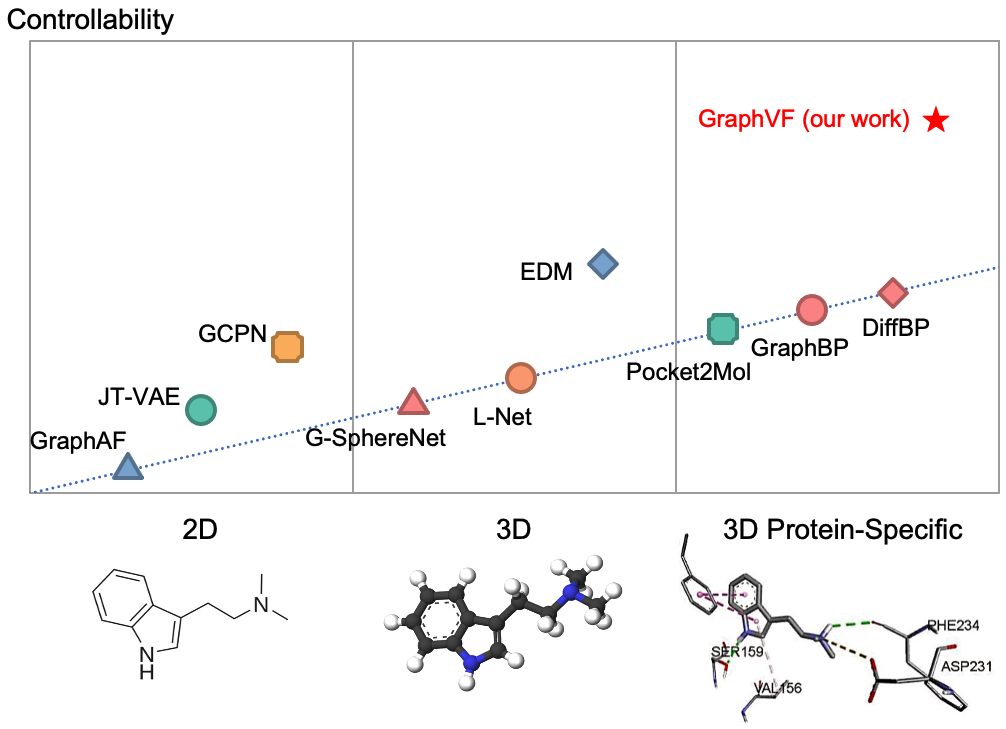}
   \caption{Overall controllability of different models across three different molecule generation tasks.}
   \label{figure:roadmap}
\end{figure}

We provide a roadmap (Figure~\ref{figure:roadmap}) to compare \model\ against previous molecule generation models. The horizontal axis indicates the three different types of tasks, whereas the vertical axis measures the overall controllability of each model in a qualitative manner. In particular, the dashed baseline represents the degree of controllability incurred by the task \textit{per se}, and the vertical distance above the dashed baseline represents the model's controllability over other specific customized properties. 

\section{Preliminaries}

\begin{figure*}[!ht]
   \centering
   \includegraphics[width=0.7\linewidth]{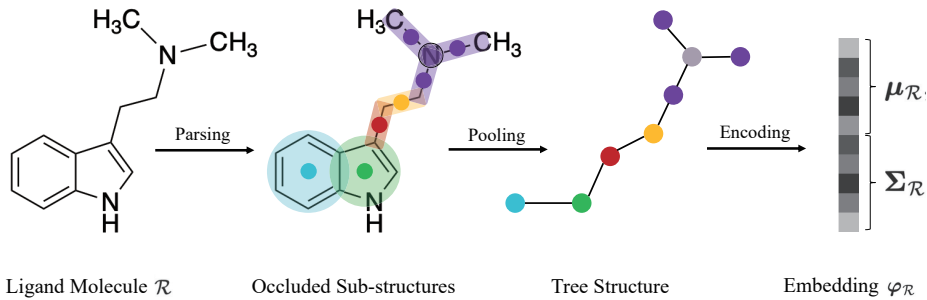}
   \caption{Ligand molecule $\mathcal{R}_{2D}$ (e.g. DMT) is first parsed into a compilation of canonical sub-structures, then pooled into a junction tree structure, and finally encoded into $\varphi_{\mathcal{R}_{2D}} = (\vmu_{\gR_{2D}}, \mSigma_{\gR_{2D})}$.}
   \label{figure:tree}
\end{figure*}

\subsection{Problem Setup}
Our goal is to generate a ligand molecule that binds effectively to a given protein receptor. 
In this paper, proteins and ligands are  represented as graphs. 
Node features in these graphs include atom type $a$ and position $r$, while edge feature involves bond type $b$. 

For training, we are given pairs of protein $\mathcal{P}$ and ligand $\mathcal{R}$ in their binding poses. 
In this paper, we denote the ligand $\mathcal{R}$'s 3D geometry graph  and  2D topology graph as $\mathcal{R}_{3D}$ and $\mathcal{R}_{2D}$, respectively. 
For generation, we are given protein targets $\mathcal{P}$ to generate drug ligands, i.e., $\mathcal{R}_{3D}$, that bind tightly to $\mathcal{P}$. 
We here consider a  protein-ligand pair with  $M$  and $N$ atoms respectively. 
Our model is trained with a set of such binding protein-ligand pairs ($\mathcal{P}$, $\mathcal{R}$). 


\subsection{Geometry Graph Encoding} \label{subsection:geo}
{3D-GNNs like SchNet~\citep{schutt2017schnet} and EGNN~\citep{satorras2021n} preserve SE(3) (\textit{i.e.} roto-translational) equivariance in the 3D space, and have been canonical in encoding 3D molecule geometry. In particular, SchNet solely relies on the relative distance between nodes during message-passing and has been efficient in modeling large bio-molecular systems like protein-ligand interaction. Specifically, geometries $\mathcal{P}$ and $\mathcal{R}_{3D}$ are organized into a radius graph, based on the Euclidean distances between atoms in the 3D space. However, this purely distance-based approach has been inadequate for modeling covalent bonds in molecular structures. Bond lengths are known to be characteristic, e.g., C$\equiv$N\ 1.16\ \r{A}, C=C 1.34\ \r{A}~\citep{lide2012characteristic}. Therefore, we explicitly incorporate bond types during massage-passing to better delineate the molecular structure and atomic interactions. We devise \textbf{EchNet}, an adapted version of SchNet, to achieve this end: } 
\begin{align}
    \mathbf{h}^{(0)}_i &= \mathrm{Emb}(a_i)\\
    \mathbf{m}_{ij} &= \mathrm{concat}\left\{\mathrm{Erbf}(||r_i - r_j||), \mathrm{Emb}(b_{ij})\right\}\label{eq:echnet-concat}\\
    \mathbf{h}^{(l)}_i &= \mathbf{h}^{(l-1)}_i + \sum_{k \in N(i) \backslash j} \mathbf{h}^{(l-1)}_k \odot \Phi^{(l)}(\mathbf{m}_{ki}),\quad l = 1,...,L
\end{align}
where $\mathrm{Erbf}(\cdot)$ is a radial 
basis function~\citep{liu2022generating}, $\mathrm{Emb(\cdot)}$ is the embedding 
layer,  $\mathrm{concat(\cdot)}$ is the concatenation of two vectors, and $\Phi$ 
is the feed-forward neural network. $L$ is the number of convolution layers, $b_{ij}$ is the bond type between atoms $i$ and $j$ (bond type `0' for non-existent bonds), 
$\mathbf{h}^{(l)}_i$ is the encoding of atom $i$ at the $l^\mathrm{th}$ 
convolution layer, and $\mathbf{m}_{ij}$ is the message propagated from atom $i$ to $j$. 
{The major difference between EchNet and SchNet resides in Equation~\ref{eq:echnet-concat}, where an extra bond-type embedding is concatenated along the distance encoding. }

\subsection{Generative Flow Model} 
\label{autoPre}
Flow-based deep generative models (\textit{i.e.},  normalizing flows)   transform a simple prior distribution into a complex data distribution by applying a sequence of invertible transformation functions. Through sampling from the prior,  new data points (i.e., new $\mathcal{R}_{3D}$ graphs) are then generated. 

To be specific, given a prior distribution $p_{Z}$, a flow model~\citep{dinh2014nice,rezende2015variational,weng2018flow} 
is defined as an invertible parameterized function $f_\theta:\mathbf{z}\in \mathbb{R}^D \rightarrow \mathbf{x}\in \mathbb{R}^D$, 
where $\theta$ represents the parameters of $f$, and $D$ is the dimension for $\mathbf{z}$ and $\mathbf{x}$.
This maps the latent variable $\mathbf{z} \sim p_{Z}$ to the data variable $\mathbf{x}$, and  the log-likelihood of  $\mathbf{x}$ is calculated as
\begin{equation}
\label{eq:flow_basic}
    \log p_{X}(\mathbf{x}) = \log p_{Z}\left(f_\theta^{-1}(\mathbf{x})\right) + \log \left|\text{det}\frac{\partial f_\theta^{-1}(\mathbf{x})}{\partial \mathbf{x}}\right|.
\end{equation}
To effectively solve the above equation, autoregressive flow model~\citep{papamakarios2017masked} formulates a flow function 
with an autoregressive computation to enable easy Jacobian determinant computation. Specifically, let $\mathbf{x}_i$ be the $i$-th component of $\mathbf{x}$ and  $\mathbf{x}_i$ 
conditions on $\mathbf{x}_{1...i-1}$. The inverse function $f_\theta^{-1}$ is 
then defined as follows: 
\begin{equation}
\label{eq:flow-paradigm}
    \mathbf{x}_i = \sigma_i(\mathbf{x}_{1...i-1})\odot \mathbf{z}_i + \mu_i(\mathbf{x}_{1...i-1}), \quad i=1...D,
\end{equation}
where $\odot$ denotes element-wise multiplication, $\sigma_i(\cdot) \in \mathbb{R}$ 
and $\mu_i(\cdot) \in \mathbb{R}$ are non-linear functions of $\mathbf{x}_{1...i-1}$. 
By doing so, we can effectively calculate the following to compute the log-likelihood in Equation~\ref{eq:flow_basic}:
\begin{equation}
\label{eq:flow-derivative}
    \mathbf{z}_i=\frac{\mathbf{x}_i - \mu_i}{\sigma_i}, \quad
    \text{det}\frac{\partial f_\theta^{-1}(\mathbf{x})}{\partial \mathbf{x}}=\prod_{i=1}^D\frac{1}{\sigma_i}.
\end{equation}
\section{The Proposed Method}

Our \model{} approach  aims to generate 3D binding molecules, namely generating $\mathcal{R}_{3D}$ that binds to $\mathcal{P}$, through sampling from a prior distribution with a variational flow model. The flow model here  aims at transforming a simple \textit{prior} distribution into the complex distribution of the 3D binding molecules by applying a sequence of invertible transformation functions as discussed in Section~\ref{autoPre}. 
As illustrated in \textbf{Figure~\ref{figure:overview}}, to  obtain desirable topological information in the generated 3D molecules ($\mathcal{R}_{3D}$), as the aim of this paper, the \model\ method encodes  the topology  patterns in 2D molecules ($\mathcal{R}_{2D}$) into the \textit{prior} distribution. 
By doing so, we can purposefully control the \textit{prior} parameterization for different generation tasks, and sampling from such \textit{prior} thus facilitates the generation of 3D molecules with specific sub-structures and physio-chemical properties encoded in the \textit{prior} distribution. 

Next, we will introduce our 2D topology prior encoding  and the 3D binding ligand generation  components in Sections~\ref{encodingsec} and~\ref{generationsec},  respectively.


\begin{figure*}[!ht]
\centering
\includegraphics[width=0.80\linewidth]{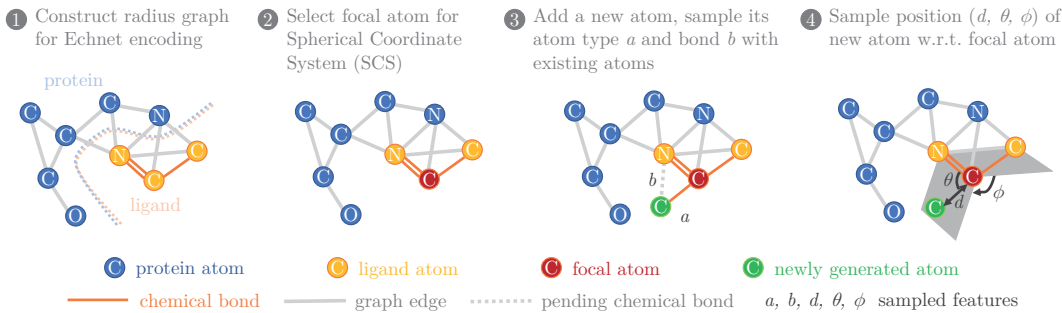}
\caption{Generation procedure of \model. Atoms are added autoregressively, whose types, 
bonds and positions are sampled from prior distribution $\mathcal N(\vmu_\mathcal{R}, \mSigma_\mathcal{R})$ 
and predicted via normalizing flow.}
\label{figure:one-step-generation}
\end{figure*}

\subsection{2D Topology Prior Encoding}
\label{encodingsec}

To encode the 2D topology in $\mathcal{R}_{2D}$ into a prior distribution, we adopt the junction tree encoder architecture from JT-VAE~\cite{jin2018junction}. 
The whole procedure is illustrated in Figure~\ref{figure:tree} and detailed next. 

\subsubsection{Ligand Scaffold Extraction and Encoding}
Following~\citep{jin2018junction}, we extract the coarse-grained structural patterns of the ligand scaffold in a fragment-driven approach. 
First, the ligand molecule $\mathcal{R}_{2D}$ (left in Figure~\ref{figure:tree}) is parsed into a compilation of occluded 
canonical sub-structures, according to a set of pre-defined vocabulary rules (detailed in Appendix~\ref{app:tree_vocab}). 
Each of such sub-structure in $\mathcal{R}_{2D}$ is then pooled into a node, resulting in a junction tree (the two middle   subfigures). 
Next, the information of this junction tree is aggregated through a  Gated Recurrent Unit (GRU)~\citep{chung2014empirical} adapted 
for tree message passing (Section~\ref{treemessagese}). 
This results in  a root node embedding $\mathbf{h}_\mathrm{root}$ representing the whole junction tree and thus the $\mathcal{R}_{2D}$.  
Finally, this embedding is then passed through a MLP to define the mean and variance of the topology distribution (right in Figure~\ref{figure:tree}): 
\begin{equation}
\label{eq:var-parameterization}
    (\vmu_{\gR_{2D}}, \mSigma_{\gR_{2D}}) = \varphi_{\mathcal{R}_{2D}} = \mathrm{MLP}(\mathbf{h}_\mathrm{root}).
\end{equation}
{Since $\vmu_{\gR_{2D}}$ and $\mSigma_{\gR_{2D}}$ are equally-sized 
dense vectors, the covariate matrix of the resultant Gaussian $\mathcal{N}(\vmu_{\gR_{2D}}, \mSigma_{\gR_{2D}})$ is diagonal. This allows us to independently sample from the different components of the Gaussian distribution (elaborated in Appendix~\ref{app:latent-encoding}).}
We go on to describe how \model{} conducts tree message passing through GRU in the junction tree to obtain the above $\mathbf{h}_\mathrm{root}$.

\subsubsection{Junction Tree Message Passing}
\label{treemessagese}
The tree message passing scheme arbitrarily selects a leaf node as the root (denoted as $\mathbf{h}_\mathrm{root}$), and passes messages from child nodes to parent nodes iteratively in a bottom-up approach. We denote the message from node $i$ to $j$ as $\mathbf{m}_{ij}$, which is updated via a GRU adapted for tree propagation:
\begin{equation}
    \mathbf{m}_{i j}=\mathrm{GRU}(\mathbf{x}_i,\left\{\mathbf{m}_{k i}\right\}_{k \in N(i) \backslash j}).
\end{equation}
To be more specific, the GRU architecture is formulated as follows:
\begin{eqnarray}
    \mathbf{s}_{ij} &=& \sum\nolimits_{k \in N(i) \backslash j} \mathbf{m}_{ki}, \\
    \mathbf z_{ij} &=& \sigma(\mathbf W^z \mathbf x_i + \mathbf U^z \mathbf{s}_{ij} + \mathbf{b}^z), \\
    \mathbf{r}_{ki} &=& \sigma(\mathbf W^r \mathbf x_i + \mathbf U^r \mathbf{m}_{ki} + \mathbf{b}^r), \\
    \widetilde{\mathbf{m}}_{ij} &=& \tanh(\mathbf W \mathbf x_i + \mathbf U \sum_{k \in N(i) \backslash j} \mathbf{r}_{ki} \odot \mathbf{m}_{ki}), \\
    \mathbf{m}_{ij} &=& (1 - \mathbf z_{ij}) \odot \mathbf{s}_{ij} + \mathbf z_{ij} \odot \widetilde{\mathbf{m}}_{ij},
\end{eqnarray}
where $\mathbf{x}_i$ is a one-hot vector, indicating the type of canonical sub-structure of node $i$. The latent representation of each node $\mathbf{h}_i$ can be derived by aggregating all the inwards messages from its child nodes as follows:
\begin{equation}
    \mathbf{h}_i = \mathbf{W}^o \mathbf{x}_i+\sum_{k \in N(i)} \mathbf{U}^o \mathbf{m}_{k i}.
\end{equation}

\subsubsection{Topology Prior}
\textbf{During training}, the 2D scaffold encoding $\varphi_{\mathcal{R}_{2D}} = (\vmu_{\gR_{2D}}, \mSigma_{\gR_{2D}})$ is regularized by a KL divergence to form 
{a compact family of diagonal Gaussians} 
around the standard gaussian $\mathcal{N}(\mathbf{0}, \mathcal{I})$. Therefore, \textbf{during generation}, we can easily navigate along 
{this family of Gaussians} 
to generate 3D molecules with specific sub-structures and physio-chemical properties encoded in the \textit{prior} distribution. 
Next, we will discuss how the flow model leverages this prior to generate binding 3D ligands. 

\subsection{3D Ligand Generation via Variational Flow}
\label{generationsec}
In this section, we will discuss how the \model{} model generates a 3D ligand $\mathcal R$ (to simplify the equations in this section, we omit the subscript), including its atoms, bonds, and geometric structure.

\subsubsection{Autoregressive Generation via Flow}
We formulate the procedure of generating a new ligand $\mathcal R$ in \model{} as a Markovian sampling process, 
where atoms and bonds are autoregressively added according to the intermediary state 
at the binding site. The generation process at step $i = 4$ is illustrated in 
Figure~\ref{figure:one-step-generation} and  
 elucidated next. 

Firstly, we construct radius graph $\mathcal{G}_i$ based on protein graph $\mathcal{P}$ 
and ligand sub-graph $\mathcal R_{1:i-1}$:
\begin{equation}
    \mathcal G_i = \tau(\mathcal P \cup \mathcal R_{1:i-1}), 
\end{equation}
where the radius operator $\tau(\cdot)$ adds edges (of bond order 0) to neighboring 
atoms within radius $\tau$. In particular, at generation step 1, when no ligand 
atoms have yet been generated, $\mathcal G_i$ is simply $\tau(\mathcal P)$. The Echnet encoder, as discussed in Section~\ref{subsection:geo}, 
outputs the encoding of each atom in both protein and ligand:
\begin{equation}
    \mathbf{\tilde e}_{1:M},\mathbf{e}_{1:i-1} = \mathrm{Echnet}(\mathcal G_i). \label{eq:encode-subgraph}
\end{equation}

{Secondly, we randomly sample a focal atom $f_i$ from all possible candidates. For each atom, its eligibility as a focal atom is determined by a binary focal classifier. Except in the first step, only atoms from the ligand molecule are considered.} 
Based on $f_i$ and two 
of its nearest neighbors, we construct a spherical coordinate system (SCS), 
transforming Cartesian coordinates into polar coordinates $(d, \theta, \phi)$. 
{Autoregressive generation under the SCS preserves the equivariance quality of our model. Refer to Appendices \ref{app:focal-atom-selection} and \ref{app:construction-scs} for implementation details of focal atom classification and SCS construction. Proof for equivariance can be found in Appendix~\ref{app:equivariance}. }

Finally, we add 
a new atom to the drug ligand via sequential generation of its 
atom type $a_i$, bindings with existing atoms $b_{1:i-1, i}$ and 
{SCS coordinates} $\mathbf{x}_i^\mathrm{(pos)} = (d_i, \theta_i, \phi_i)$, in order to better capture the underlying 
dependencies \citep{liu2022generating}. 
{
This is achieved by sampling the prior random variables 
$\mathbf{z}_i^\mathrm{(node)}$, $\mathbf{z}_{1:i-1, i}^\mathrm{(bond)}$ 
and $\mathbf{z}_i^\mathrm{(pos)}$ from the variational distribution $\mathcal N(\vmu_{\gR}, \mSigma_{\gR})$:} 
\begin{equation}
\left[\mathbf{z}_i^{\text {(node)}} ; \mathbf{z}_{1: i-1, i}^{\text {(bond)}} ; \mathbf{z}_i^{\text {(pos)}}\right] \sim \mathcal{N}\left(\boldsymbol{\mu}_{\mathcal{R}}, \boldsymbol{\Sigma}_{\mathcal{R}}\right). 
\end{equation}
{Recall from Equation~\ref{eq:var-parameterization} that $\mathcal{N}\left(\boldsymbol{\mu}_{\mathcal{R}}, \boldsymbol{\Sigma}_{\mathcal{R}}\right)$ is parameterized as a diagonal Gaussian, so each component of the random variable can be sampled independently from each other. In particular, $\mathbf{z}^{\text {(bond)}}$ is repeatedly sampled for $(i-1)$ times because we need to determine the bond type between the new atom $i$ and each of the previous $(i-1)$ ligand atoms.}
The priors are then consecutively projected 
to the 3D geometric space via flow transformation $\mathcal{F}_i$:
\begin{equation}
    \mathbf{x}_i^\mathrm{(node)}, \mathbf{x}_{1:i-1, i}^\mathrm{(bond)},\ \mathbf{x}_i^\mathrm{(pos)} = \mathcal{F}_i\left(\mathbf{z}_i^\mathrm{(node)}, \mathbf{z}_{1:i-1, i}^\mathrm{(bond)},\ \mathbf{z}_i^\mathrm{(pos)} ;\mathbf{e}_{1:i-1}\right).
    \label{eq:flow_forward}
\end{equation}
\subsubsection{Parameterization of the Flow Transformation}
{Following the paradigm described in Equatioin~\ref{eq:flow-paradigm}, }the above flow transformation $\mathcal{F}_i$ is parameterized with {the subsequent steps}:
\begin{align}
    \mu_i^\mathrm{(node)}, \sigma_i^\mathrm{(node)} &= \text{Node-MLP}(\mathbf{e}_{f_i})\label{eq:node_mlp},\\
    \mathbf{x}_i^\mathrm{(node)} &= \sigma_i^\mathrm{(node)} \odot \mathbf{z}_i^\mathrm{(node)} + \mu_i^\mathrm{(node)},\\
    \mu_{1:i-1, i}^\mathrm{(bond)}, \sigma_{1:i-1, i}^\mathrm{(bond)} &= \text{Bond-MLP}(\mathbf{e}_{1:i-1}, \mathbf{x}_i^\mathrm{(node)})\label{eq:bond_mlp},\\
    \mathbf{x}_{1:i-1, i}^\mathrm{(bond)} &= \sigma_{1:i-1, i}^\mathrm{(bond)} \odot \mathbf{z}_{1:i-1, i}^\mathrm{(bond)} + \mu_{1:i-1, i}^\mathrm{(bond)},\\
    \mu_i^\mathrm{(pos)}, \sigma_i^\mathrm{(pos)} &= \text{Position-MLP}(\mathbf{e}_{f_i}, \mathbf{x}_i^\mathrm{(node)}, \mathbf{x}_{1:i-1, i}^\mathrm{(bond)})\label{eq:position_mlp},\\
    \mathbf{x}_i^\mathrm{(pos)} &= \sigma_i^\mathrm{(pos)} \odot \mathbf{z}_i^\mathrm{(pos)} + \mu_i^\mathrm{(pos)}, \label{eq:pos-flow}
\end{align}
{where $\mathbf{e}_{f_i}$ is the encoding of the focal atom $f_i$, and }
{Node-MLP, Bond-MLP, and Position-MLP are layers of flow MLPs, detailed in Appendix~\ref{app:implementation}.}   
$\odot$ denotes element-wise multiplication, and $\mathbf{x}_i^\mathrm{(node)}, \mathbf{x}_{1:i-1, i}^\mathrm{(bond)}$, $\ \mathbf{x}_i^\mathrm{(pos)} $ 
are the vectorized representation of atom type, bond type, and SCS-based position, and 
$\sigma, \mu$ are parameters for flow transformation. The sequential dependencies 
between $a, b, d, \theta, \phi$ are embodied in Equations~\ref{eq:bond_mlp} and~\ref{eq:position_mlp}, 
where new atom/bond types that have just been generated are immediately used to 
parameterize $\sigma$ and $\mu$ of the next flow transformation.

Thus, we have rendered all the sampled features $a_i, b_{1:i-1, i}, d_i, \theta_i$, $\phi_i$ 
from step $i$, and successfully generate the new atom and its associated bonds. 
We go on with this iteration, until the focal classifier reports that no atom 
is eligible for $f_i$, and the generation procedure is called to an end. 
Algorithm 2 from Appendix~\ref{app::algorithm} explains the 
generation algorithm in more detail.

\subsubsection{The Objective Function} 
\label{subsection:varflow}

As discussed in Section~\ref{autoPre}, the generative flow model explicitly learns the data distribution $p (x)$  and thus has the simple loss  function as the negative log-likelihood (denoted as  $\mathcal{L}_\mathrm{flow}$). We here add a regularization term (denoted as  $\mathcal{L}_\mathrm{KL}$) to align the topology prior as discussed in Section~\ref{encodingsec} with a standard Gaussian. This results in the following loss for GraphVF:
\begin{equation}
    \mathcal{L}_\mathrm{total} = \mathcal{L}_\mathrm{flow} + \beta \mathcal{L}_\mathrm{KL},
\end{equation}
where $\beta$ is a hyper-parameter that trades off between the two loss terms. 

To gain more intuition into the construction of this loss function, we formulate the goal of variational training into a bi-level optimization task: 
\begin{align}
    \mathrm{maximize\quad} & p_{X_i}(\rvx_i)\ ,\label{eq:maximize_p}\\
    \mathrm{subject\ to\quad} & D_\mathrm{KL}(X_i ||\hat X_i) < \xi,  \label{eq:subject_KL} 
\end{align} 
where {$ \xi$ is the tolerance threshold for distribution shift between the generated ligands and the reference ligands}. 
Since $\mathcal{F}_i$ is an invertible function ($i = 1...N$), the LHS of Inequality~\ref{eq:subject_KL} can be transformed into:
\begin{align}
    \mathcal{L}_\mathrm{KL}^b
    & = D_\mathrm{KL}(X_i ||\hat X_i) \\
    &= D_\mathrm{KL}(\mathcal{F}_i(Z) || \mathcal{F}_i(\hat Z) ) \\
    &= D_\mathrm{KL}(Z ||\hat Z)\\
    &= D_\mathrm{KL}(\mathcal{N}(\vmu_\gR, \mSigma_\gR) || \mathcal{N}(\boldsymbol{0}, \boldsymbol{I})),
\end{align}
where $b$ stands for a particular protein-ligand pair. 

As for Equation~\ref{eq:maximize_p}, it corresponds to the flow loss term described 
in Equation~\ref{eq:flow_basic}. 
In the flow-based model, the training process is the exact inverse of the generation process: 
\begin{equation}
\mathbf{z}_i^\mathrm{(node)}, \mathbf{z}_{1:i-1, i}^\mathrm{(bond)},\ \mathbf{z}_i^\mathrm{(pos)} = \mathcal{F}_i^{-1}\left(\mathbf{x}_i^\mathrm{(node)}, \mathbf{x}_{1:i-1, i}^\mathrm{(bond)},\ \mathbf{x}_i^\mathrm{(pos)} ;\mathbf{e}_{1:i-1}\right).
\end{equation}
To be specific, $\mathbf{z}_i$ is derived as follows in the training phase:
\begin{align}
\label{eq:flow_train_1}
\mathbf{z}_i^\mathrm{(node)} &= \left(\mathbf{x}_i^\mathrm{(node)} - \mu_i^\mathrm{(node)}\right) \odot \frac{1}{\sigma_i^\mathrm{(node)}}\\
\label{eq:flow_train_2}
\mathbf{z}_{1:i-1, i}^\mathrm{(bond)} &= \left(\mathbf{x}_{1:i-1, i}^\mathrm{(bond)} - \mu_{1:i-1, i}^\mathrm{(bond)}\right) \odot \frac{1}{\sigma_{1:i-1, i}^\mathrm{(bond)}}\\
\label{eq:flow_train_3}
\mathbf{z}_i^\mathrm{(pos)} &= \left(\mathbf{x}_i^\mathrm{(pos)} - \mu_i^\mathrm{(pos)}\right) \odot \frac{1}{\sigma_i^\mathrm{(pos)}}, 
\end{align}
where $\sigma_i$ and $\mu_i$ are derived in the same way as Equations~\ref{eq:node_mlp}, 
\ref{eq:bond_mlp} and~\ref{eq:position_mlp}.

For step $i$ in 
{a given drug-ligand} pair $b$, we substitute equations \ref{eq:flow_train_1}$\sim$\ref{eq:flow_train_3} into 
{Equations \ref{eq:flow_basic} and \ref{eq:flow-derivative} to get flow loss term:} 
\begin{align}
\mathcal{L}_i^\mathrm{(node)}  &= -\log(\operatorname{Prod}(\mathcal N(\mathbf{z}_i^\mathrm{(node)}|\vmu_\gR, \mSigma_\gR))) -\log(\operatorname{Prod}(\frac{1}{\sigma_i^\mathrm{(node)}})) \\
\mathcal{L}_{1:i-1, i}^\mathrm{(bond)}  &= -\log(\operatorname{Prod}(\mathcal N(\mathbf{z}_{1:i-1, i}^\mathrm{(bond)}|\vmu_\gR, \mSigma_\gR))) -\log(\operatorname{Prod}(\frac{1}{\sigma_{1:i-1, i}^\mathrm{(bond)}})) \\
\mathcal{L}_i^\mathrm{(pos)}  &= -\log(\operatorname{Prod}(\mathcal N(\mathbf{z}_i^\mathrm{(pos)}|\vmu_\gR, \mSigma_\gR))) -\log(\operatorname{Prod}(\frac{1}{\sigma_i^\mathrm{(pos)}})) \\
\mathcal{L}^{i, b}_\mathrm{flow} &= \mathcal{L}_i^\mathrm{(node)} + \mathcal{L}_{1:i-1, i}^\mathrm{(bond)} + \mathcal{L}_i^\mathrm{(pos)}. 
\end{align}

{We further take the binary focal classifier into consideration. For step $i$ of pair $b$, the flow loss term is formulated as: } 
\begin{equation}
\mathcal{L}_\mathrm{focal}^{i, b} = 
\begin{cases}
\frac{1}{M} \sum_{j = 1}^M \mathrm{BCELoss}(\tilde{y}_j, \Hat{\tilde{y}}_j), & \text{if\ } i = 1; \\
\frac{1}{i-1} \sum_{j = 1}^{i-1} \mathrm{BCELoss}({y}_j, \Hat{{y}}_j), & \text{if\ } i > 1,  \\
\end{cases}
\end{equation}
{where $\mathrm{BCELoss(\cdot)}$ is the binary cross entropy, $y_j$ is the predicted focal score of atom $j$, and $\hat{y}_j$ is its ground-truth label. $\tilde{y}_j$ stands for the $j$-th atom from the protein receptor. The ground truth label for focal atom classification during training is curated from a ring-first expert trajectory~\citep{li2021learning}, which preserves ring locality and boosts model performance (Appendix~\ref{app:ring-first-traversal}). }

Therefore, the $\beta$-regularized loss form can be finalized as 
\begin{equation}
    \mathcal{L}^{b}_\mathrm{total} = \frac{1}{N}\sum_{i=1}^{N}(\mathcal{L}^{i, b}_\mathrm{flow} + \mathcal{L}^{i, b}_\mathrm{focal}) + \beta \mathcal{L}_\mathrm{KL}^b.  
\end{equation}
Algorithm 1 from Appendix~\ref{app::algorithm} 
explains the training algorithm in more detail.

\subsection{Controllable generation } 
\label{subsubsection:control_gen} 
During generation, the variational prior in \model{} provides a flexible interface for controlling 
certain properties of the generated 3D molecules in a zero-shot manner, without the 
need to re-train the model. For a certain desired ligand attribute $\rho$, there are 
two ways to acquire its corresponding prior: 

\paragraph{1) Mean aggregation. } This is suitable for qualitative attributes, such as the existence of a certain pharmacophore or sub-structure. We collect a set of such molecules with attribute $\rho$ (denoted as $\{\gR_{a}\}_{a\in I}$, where $I$ is the index set), and carry out mean aggregation over their structural encodings: 
\begin{equation}
    (\vmu_\rho, \mSigma_\rho) = \frac{1}{|I|} \sum_{a\in I}(\vmu_{\gR_a}, \mSigma_{\gR_a}). 
\end{equation}

\paragraph{2) Bayesian Optimization. } This approach is ideal for numerical attributes like free energy and enthalpy. We use a sparse gaussian process (SGP) to fit the relationship $f$ between latent encoding and the desired property value 
\begin{equation}
    f: (\vmu_{\gR_a}, \mSigma_{\gR_a}) \mapsto ||\rho||. 
\end{equation}
upon this relation $f$, we perform bayesian optimization to find the latent encoding that corresponds to the maximum value of $||\rho||$: 
\begin{equation}
    (\vmu_\rho, \mSigma_\rho) = \mathop{\mathrm{argmax}}_{\vmu, \mSigma} f(\vmu, \mSigma).
\end{equation}



\begin{table*}[h]
    
    \begin{center}
    \begin{tabular}{ccccccccc}
    \toprule
    \multicolumn{1}{c}{\bf Method} &
    \multicolumn{1}{c}{\bf HA$\uparrow$} &
    \multicolumn{1}{c}{\bf SA$\uparrow$} &
    \multicolumn{1}{c}{\bf Lipinski$\uparrow$} &
    \multicolumn{1}{c}{\bf QED$\uparrow$} &
    \multicolumn{1}{c}{\bf LogP} &
    \multicolumn{1}{c}{\bf Novelty$\uparrow$} &
    \multicolumn{1}{c}{\bf Diversity$\uparrow$} &
    \multicolumn{1}{c}{\bf Time (s)$\downarrow$}
    \\ \midrule
    GraphBP & 0.134 & 0.585 & 4.909 & 0.515 & 2.610 & 0.569 & 0.835 & \textbf{20}\\
    Pocket2Mol & 0.272 & \textbf{0.765} & 4.920 & \textbf{0.563} & 1.586 & 0.624 & 0.688 & 2504\\
    GraphVF (w/o 2D encoder)& 0.263 & 0.542 & 4.080 & 0.310 & 3.567 & 0.605 & 0.807 & 68\\
    GraphVF (ours)&\textbf{0.311} & 0.585 & \textbf{4.968} & 0.448 & 0.699 & \textbf{0.737} & \textbf{0.930} & 32\\
    \bottomrule
    \end{tabular}
    \end{center}
    \caption{Performance of different methods on 3D molecular generation based 
    on protein pockets. Best results are in \textbf{bold}. 
    }
    \label{table:main-evaluation-results}
\end{table*}

\begin{figure*}[!ht]
\centering
\includegraphics[width=0.9\textwidth]{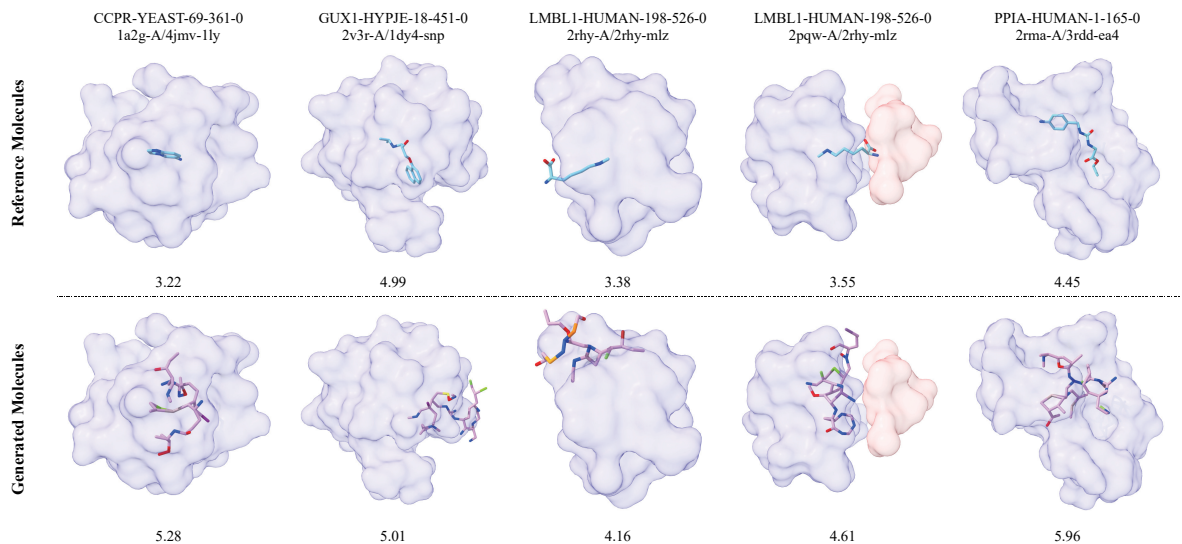}
\caption{Examples of generated molecules with higher binding affinity (Gnina score$\ \uparrow$) than the reference molecules. Protein names and residue/ligand IDs are listed on top. We use ChimeraX~\citep{goddard2018ucsf} for visualization. 
}
\label{figure:surface}
\end{figure*}

\section{Experiments}
We conduct three sets of experiments to verify the effectiveness of our \model{} model: 

\begin{itemize}
    \item Basic protein-specific generation. Prior set to $\mathcal{N}(\mathbf{0}, \mathcal{I})$. 
    \item Protein-specific generation with controlled sub-structure. Prior selected using mean aggregation. 
    \item Protein-specific generation with controlled properties. Prior selected using Bayesian Optimization. 
\end{itemize}

\subsection{3D Molecular Generation Conditioned on Protein Pocket}

\textbf{Dataset.} We use the benchmarking CrossDocked dataset~\citep{francoeur2020three}, 
which contains 22.5 million protein-ligand pairs, to evaluate the generation 
performance of GraphVF. For fair comparison, we follow Pocket2Mol~\citep{peng2022pocket2mol} 
to prepare and split the data.

\begin{table*}[!ht]
\begin{center}
\begin{tabular}{ccccc}
\toprule
\multicolumn{1}{c}{\bf Sub-Structure} &
\multicolumn{1}{c}{\bf GraphBP} &
\multicolumn{1}{c}{\bf Pocket2Mol} &
\multicolumn{1}{c}{\bf GraphVF (w/o 2D encoder)} &
\multicolumn{1}{c}{\bf GraphVF}
\\ \midrule
CC & 0.27 & 2.18 & 1.05 & \textbf{0.22} \\
C=O & 0.83 & 3.78 & 0.73 & \textbf{0.67} \\
CN & \textbf{0.70} & 1.78 & 1.27 & 0.77 \\
\hline
CCC & 1.31 & \textbf{1.02} & 1.07 & 1.25 \\
CCO & 0.61 & 1.38 & 0.52 & \textbf{0.51} \\
NCC & 1.00 & 1.05 & \textbf{0.62} & 0.83 \\
\hline
CCCC & 2.15 & 2.10 & 2.03 & \textbf{2.00} \\
CCCO & 2.37 & 2.27 & 2.17 & \textbf{2.17} \\
CC=CC & 2.20 & 2.85 & 2.70 & \textbf{2.04} \\
\bottomrule
\end{tabular}
\end{center}
\caption{The KL divergence of the distance (upper part), bond angles (middle part) and dihedral angles (lower part) with the test set. The best results are in \textbf{bold}.}
\label{table:sub-structure-analysis}
\end{table*}

\textbf{Setup.} Following GraphBP~\citep{liu2022generating} and Pocket2Mol, we 
randomly sample 100 molecules for every protein pocket in the generation stage. 
The quality of generated molecules is evaluated by 8 widely adopted metrics: 
\begin{enumerate}
    \item \textbf{ High Affinity (HA)}, which estimates the percentage of generated molecules that have higher \emph{CNNAfinity} calculated by the \emph{Gnina} program~\citep{mcnutt2021gnina}. Gnina score has been demonstrated to be consistently more accurate and un-biased than Vina score in previous works~\citep{liu2022generating, peng2023pocketspecific}. For this reason, we do not report Vina-based HA;
    \item \textbf{ Synthetic Accessibility (SA)}, 
which represents the easiness of drug synthesis; 
    \item \textbf{ Lipinski} estimates the mean number of Lipinski rules followed by the generated molecules; 
    \item \textbf{ Quantitative 
Estimation of Drug-likeness (QED)}, a measure of drug-likeness based on desirability~\citep{bickerton2012qed};
    \item \textbf{ LopP} denotes 
the octanol-water partition coefficient. Good drug candidates have LogP between -0.4 and 5.6~\citep{ghose1998logp}; 
    \item \textbf{Novelty} is calculated as $1$ $-$ the average of maximum Tanimoto similarities to training set molecules among the generated molecules;
    \item \textbf{Diversity} is calculated as $1$ $-$ the average Tanimoto similarities of generated molecules for every protein pockets, following Pocket2Mol;
    \item \textbf{Time} estimates the time(s) spent on generating 100 molecules for a pocket.
\end{enumerate}
We choose GraphBP and Pocket2Mol as our baselines, which represent the state-of-art models for binding molecule generation. For GraphBP and GraphVF, we trained them on the dataset for 40 epochs with the same hyperparameters. For Pocket2Mol ~\citep{peng2022pocket2mol}, we obtain the pre-trained model from their authors and then compute the scores using \emph{Gnina}.

\textbf{Results.} 
The comparison results are presented in Table~\ref{table:main-evaluation-results}. 
We can see that our \model\ outperforms the two strong baselines in 
terms of HA, Lipinski, LogP, Novelty, and Diversity. As is shown by the generation time, \model{} is much more efficient than Pocket2Mol in 2 orders of magnitude. 
We have also attained reasonably good performance on drug properties like QED, 
LogP and SA, even without explicit guidance from the variational prior. This shows that our model learns good molecular representations by combining 2D topology with 3D geometry, and is able to generate robustly good molecules under a different prior. The great novelty and diversity of our molecules can be attributed to our variational training strategy. As is exemplified in Figure~\ref{figure:surface}, the distribution shift between training and generation allows \model\ to explore larger and more complex molecules in the chemical space, and can hopefully generate novel drug-like molecules that have never been discovered before. 


\textbf{{Ablation study. }} A critical question about the variational flow architecture is: Will the KL loss term (\eqref{eq:subject_KL}) collapse the variational distribution to a standard gaussian, and degrade \model\ to a purely flow-based model? This is not happening, because our model can learn potentially better molecular representations from $\varphi_{\mathcal{R}_\mathrm{2D}}$. To empirically prove this claim and mimic the degradation scenario, we perform an ablation study by masking the 2D encoder $\varphi_{\mathcal{R}_\mathrm{2D}}$ and substituting the prior as the standard gaussian $\mathcal{N}(\mathbf{0}, \mathcal{I})$ during training. From the 3rd row of Table~\ref{table:main-evaluation-results}, we can see that the ablated `GraphVF (w/o 2D encoder)' version performs much worse. In particular, the HA value for the ablated variant drops drastically from 31.1\% to 26.3\%. This shows that the variational-flow framework is very effective in balancing the 2D/3D dual data sources and preventing model degradation. 

\subsection{Sub-structure Analysis}

\textbf{Setup.} As it is pointed out in Pocket2Mol~\cite{peng2022pocket2mol} 
that conventional metrics could not reflect the geometry of sampled molecules, 
we conduct additional sub-structure analysis. Following Pocket2Mol, We compare GraphVF with previous works by the KL divergence between the distributions of generated bond length, bond angles, dihedral angles and the corresponding distributions of the test set.

\textbf{Results.} The results are presented in Table~\ref{table:sub-structure-analysis}. 
In comparison to GraphBP and Pocket2Mol, GraphVF yields the best results on 
dihedral angles, which indicates that it is more capable of modeling complex dependencies. 
At the same time, it achieves comparable results to GraphBP and Pocket2Mol on bond length and bond angles. Ablation study shows 2D encoder helps GraphVF better capture the geometry of sub-structures.

\begin{figure}[!ht]
   \centering
   \includegraphics[width=\linewidth]{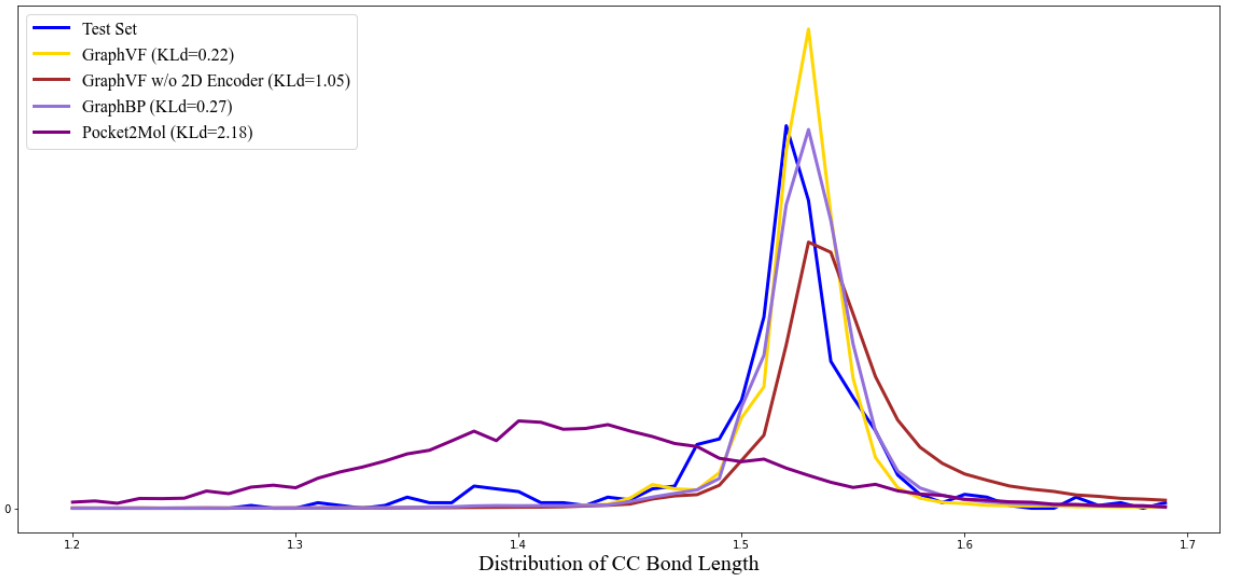}
   \includegraphics[width=\linewidth]{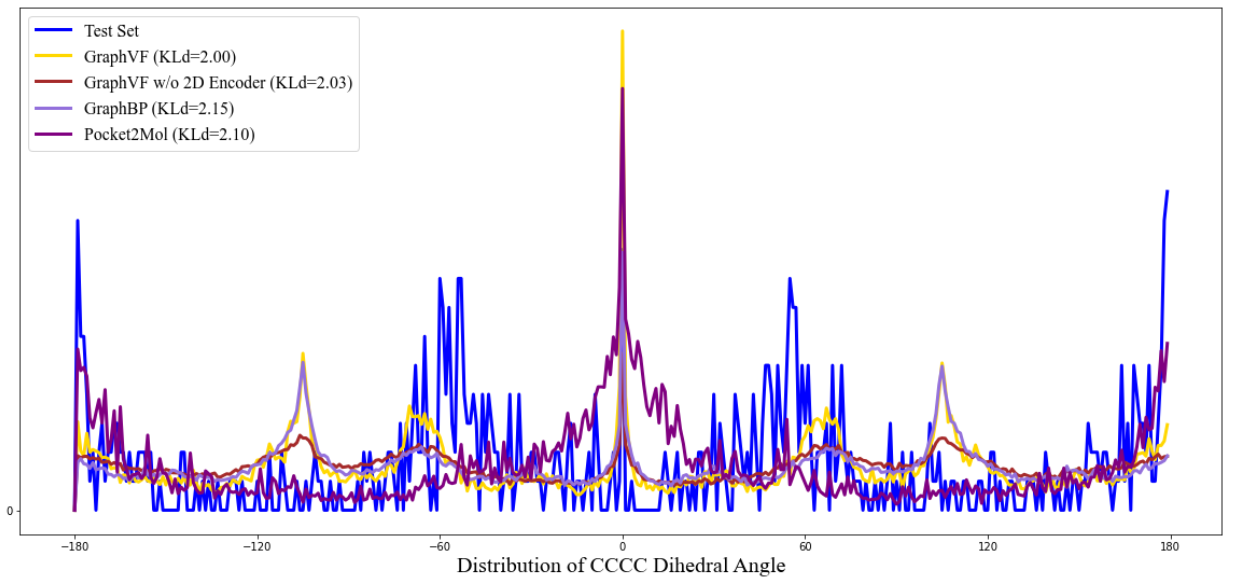}
   \caption{
   {Examples of sub-structural distributions of the test set and those generated from models. `KLd' in parentheses stands for the KL divergence between the test set distribution and the distribution of the generated data. (a) Distribution of CC bond length; (b) Distribution of CCCC dihedral angle. }}
   \label{figure:sub-structure-vis}
\end{figure}

We visualize the distribution of CC bond length and CCCC dihedral angle in Figure~\ref{figure:sub-structure-vis}. In both cases, GraphVF grasps geometry features the best so that it produces a distribution that is the closest to the test set. This shows that our emphasis on the delineation of molecular structures really pays off. The realistic sub-structural distribution is the basis for generating high-quality drug molecules and good controllability.

\begin{table*}[h]
\vspace{-0.5em}
\begin{center}
\begin{tabular}{cccccc}
\toprule
{\bf Rate of specified sub-structure(\%)} &{\bf Test Set} &{\bf GraphBP} &{\bf Pocket2Mol} &{\bf GraphVF w/ latent $\rho$} &{\bf GraphVF w/o latent $\rho$} \\
\midrule
alkenyl&76&41.8&90.7&\textbf{93.0}&\underline{66.2} \\
imine&47&5.5&51.7&\textbf{83.5}&\underline{49.7} \\
5-member ring containing element S&10&2.1&2.7&\textbf{45.0}&\underline{5.4} \\
6-member ring containing element O&13&4.7&\underline{14.3}&\textbf{86.5}&17.1 \\
\bottomrule
\end{tabular}
\end{center}
\caption{Controllable generation for specified sub-structures. The highest rates are in bold; rates closest to test set are underlined.  
}
\label{table:substructure-result}
\end{table*}

\begin{table*}[h]
\vspace{-0.5em}
\begin{center}
\begin{tabular}{cccccc}
\toprule
{\bf Average property value} &{\bf Test Set} &{\bf GraphBP} &{\bf Pocket2Mol} &{\bf GraphVF w/ latent $\rho$} &{\bf GraphVF w/o latent $\rho$} \\
\midrule
$\epsilon_{\textrm{HOMO}}/\mathrm{eV}$ & -6.64 & -7.03 & -6.78 & \textbf{-6.60} & -6.68 \\
$U_0/\mathrm{eV}$ & -174.35 & -148.29 & -141.00 & \textbf{-108.76} & -202.17 \\
$U/\mathrm{eV}$ & -175.43 & -149.03 & -141.55 & \textbf{-127.07} & -202.31 \\
$H/\mathrm{eV}$ & -176.83 & -150.28 & -142.63 & \textbf{-123.40} & -208.27 \\
$G/\mathrm{eV}$ & -160.83 & -137.24 & -128.75 & \textbf{-106.63} & -183.01 \\
\bottomrule
\end{tabular}
\end{center}
\caption{Controllable generation for specified molecular properties. The highest values are in bold. 
}
\label{table:property-result}
\end{table*}

\color{black}
\subsection{Controllable Generation for Specified Chemical Sub-structures}
\label{subsection:sub_structure}

Our pretrained framework could be used to encourage desired sub-structures during generation. We carry out case studies on generation of 
molecules containing the following motifs: alkenyl, imine, 5-member ring containing element S, and 6-member ring containing element O. For each motif, we use mean aggregation to calculate latent distribution
$\mathcal{N}(\vmu_\rho, \mSigma_\rho)$ from 500 randomly sampled reference ligand molecules 
in the training set that contain the motif as a sub-structure. Finally, we calculate the rate of the 
generated molecules that contain the specified sub-structures on the test set, which 
is compared with the results of directly sampling from prior distribution 
$\mathcal{N}(\boldsymbol{0}, \boldsymbol{I})$.

The experimental results are summarized in Table~\ref{table:substructure-result}. 
With the prior distributions collected from molecules that contain specified 
sub-structures, our model is more likely to generate ligand molecules with those 
sub-structures. When directly sampled from prior distribution $\mathcal{N}(\boldsymbol{0}, \boldsymbol{I})$, our model generates specified sub-structures at a rate which is the closest to the appearing rate of these structures in the test set, in comparison with GraphBP and Pocket2Mol.

\subsection{Controllable Molecular Generation for Specified Molecular Properties}
\label{subsection:drug_property}

Our framework can also be explicitly controlled to generate drug-like molecules 
with desired properties. To support this claim, we perform case studies to optimize the quantum-mechanical (QM) properties of the generated drug ligands right at the binding site. Since \textit{ab-initio} QM property calculations are very computationally expensive, we use the DimeNet~\citep{klicpera2020fast} model for property prediction, which has been pretrained on the QM-9~\citep{ramakrishnan2014quantum} dataset. We select 5 important physio-chemical properties for prediction: 
\begin{itemize}
    \item Highest occupied molecular orbital energy:  $\epsilon_{\textrm{HOMO}}/\mathrm{eV}$;
    \item Internal energy at 0K:  $U_0/\mathrm{eV}$; 
    \item Internal energy at 298.15 K: $U/\mathrm{eV}$; 
    \item Enthalpy at 298.15K: $U/\mathrm{eV}$; 
    \item Free energy at 298.15K: $U/\mathrm{eV}$. 
    
\end{itemize}

We retrieve these properties of our CrossDocked training-set molecules from DimeNet, and their latent encodings from the tree encoder. As described in Section~\ref{subsubsection:control_gen}, we use SGP to fit the property-encoding relationship, and then find the optimal latent encoding $\mathcal{N}(\vmu_\rho, \mSigma_\rho)$ with the \textbf{highest} energy through bayesian optimization, which is further used for controllable property generation. 

Experiment results are presented in Table~\ref{table:property-result}. Our model achieves consistently higher energy than all the baselines, which clearly shows that the latent $\varphi_\rho$ is effective in curating  desired properties of the generated molecules.

\section{Concluding remarks}
We propose \model, a novel variational flow-based framework for controllable 
binding 3D molecule generation.  We empirically demonstrate that, through 
effectively integrating  2D structure semantics and 3D pocket geometry, \model\ 
can obtain superior performance to the state-of-the-art strategies for pocket-based 
3D molecule generation. We also experimentally show that, \model\ can effectively 
generate binding molecules with desired ligand sub-structures and bio-chemical properties.  
 
Our work here demonstrates that domain constraints can be effectively leveraged 
by deep generative models to improve the qualities of molecule design and fulfill 
the need for controllable molecule generation. Our studies here shed light on 
the potential of  generating binding ligands with sophisticated domain knowledge 
and finer-grained control over a variety of bio-chemical properties. 

\newpage

\bibliographystyle{ACM-Reference-Format}
\vspace{-2.05mm}
\bibliography{reference}

\newpage
  
\newpage
\appendix

\onecolumn

\Huge{\noindent \textbf{Appendices}}

\normalsize

\section{Ligand scaffolds encoding}\label{app:tree}
\subsection{Parsing and pooling}\label{app:tree_vocab}
There are 3 types of canonical sub-structures, as exemplified by the DMT molecule in Figure~\ref{figure:tree}:
\begin{enumerate}
    \item Rings, e.g. the blue, green nodes;
    \item Non-ring covalent atom pairs, e.g. the red, yellow and purple nodes;
    \item Pivot atoms that are connected to 3 or more items, e.g. the gray node.
\end{enumerate}
The rules for identifying sub-structures are self-contained, yielding a relatively sparse and stable set of vocabulary. A total of 427 canonical sub-structures are identified from the 100,000 reference ligands in the CrossDocked dataset.
Once the ligand molecule is parsed into a compilation of sub-structures, the molecular graph can be pooled into a junction tree in a straightforward manner, where each sub-structure corresponds to a tree node, and any two intersecting sub-structures yield an edge between their corresponding nodes.

\subsection{Latent encoding}\label{app:latent-encoding}

Finally, the structural encoding of the whole molecule is obtained by feeding $\mathbf{h}_\mathrm{root}$ through a MLP: 
\begin{equation}
    (\vmu_\gR, \mSigma_\gR) = \varphi_\mathcal{R} = \mathrm{MLP}(\mathbf{h}_\mathrm{root}). 
\end{equation}
{Two equally-shaped ($34\times 1$) dense vectors $\vmu_\gR$ and $\mSigma_\gR$ comprise the latent encoding
and parameterize the mean and variance of an amortized diagonal Gaussian distribution $\mathcal{N}(\vmu_\gR, \mSigma_\gR)$. 
Their different channels serve different purposes during sampling. A de-quantized one-hot vector for atom type (27 possible choices) is sampled from the first 27 channels. A de-quantized one-hot vector for bond type (single, double, triple, or none) is sampled from the next 4 channels. The latent values for $(d, \theta, \phi)$ are directly sampled from the last three channels because they are continuous by nature. The gaussian distribution $\mathcal{N}(\vmu_\gR, \mSigma_\gR)$ is parameterized to have a diagonal co-variate matrix, so each channel of this gaussian is independent from each other.  }

All parameters in this section are \textbf{trainable}. Our earlier attempts with a pre-trained version of GRU would result in serious degradation of the quality of generated molecules. Thus, end-2-end training of these parameters is ideal for achieving better model performance.

\section{Implementation Details}\label{app:implementation}

\textbf{\quad Network Architecture.} We stack 6 layers of the Echnet and 20 layers of the tree GRU. We use 6 variational flow layers for generation.


\textbf{Training Details.} We train GraphVF for 40 epochs on the full training set with batch size 4. We use Adam optimizer while setting learning rate as 1e-4 and weight decay as 1e-6. For the $\beta$-annealing which is applied to the whole training process, we pick the minimum $\beta$ as 1e-4 and maximum $\beta$ as 0.015.

\textbf{Generation Details.} We sample 100 molecules for each pocket in the test set. Molecules that have less than 15 atoms are excluded and re-sampled, while molecules that have more than 50 atoms are truncated at the 50-th atom. Additionally, to help GraphVF generate ligand molecules with good geometric properties, we propose to limit the sample space by five validity constraints during generation:

\begin{enumerate}
    \item A bond always exists between the newly generated atom and the focal atom;
    \item At most one other atom could be connected to the newly generated atom with a bond;
    \item The newly generated atom could only have bonds with atoms that are predicted positive by the focal classifier;
    \item The element of generated atom always lies in C, N, O, P, S, Cl; 
    \item The length of all generated bonds should be less than 10 \AA; 
    \item When all the generated molecules have a diameter $>$ 10 \AA, all unfinished molecules should be dropped for a new round of generation.
\end{enumerate}

These constraints can be flexibly applied, without the need to re-train the model from scratch. Therefore, they can be duly employed to our generation process to achieve uniformly good results on different benchmarks. 


\textbf{Sub-structure Analysis.} To approximate the distribution for visualization and calculation of KL divergence, we set 0.01 \AA\ per bin for distance and 1 degree per bin for bond angles and dihedral angles. In total, we have 2,000 bins for distances ranging from 0 to 20 \AA, 180 bins for bond angles and 360 bins for dihedral angles. To avoid 0 during calculating KL divergence, we replace every 0-count with 1 / number-of-bins. For fair comparison, we use every model to sample 100 ligand molecules for each protein in the test set.

\textbf{Controllable Generation for Specified Sub-structures.} 
When we sample from the training set, 
molecules with more than 16 atoms are dropped because they are likely to be less representative owing to having multiple sub-structures.

\textbf{Controllable Generation for Specified Properties. } We use DimeNet++ pretrained on QM-9 dataset (preset as part of the DGL library) for generation. We implement Bayesian optimization using BoTorch~\citep{balandat2020botorch}, a framework for efficient Monte-Carlo Bayesian Optimization. Since SGP has $\mathcal{O}(n^3)$ time complexity and  $\mathcal{O}(n^2)$ space complexity, we can only afford to perform BO on a 1/10 training dataset with 10000 reference ligand molecules. We use an upper confidence bound (UCB) with $\beta = 0.1$. In the acquisition function, we use 512 raw samples for initializations and set 10 re-starts to get the top 1 encoding with the highest property score. During the inference stage, we notice that DimeNet can produce unrealistic property scores close to infinity, so we regularize the prediction result using the $3\sigma$ principle, where the mean and variance statistics is gleaned from the QM-9 dataset labels. This regularization effectively covers about 90\% of all the outputs. 
 
\newpage
\section{Algorithms for training and generation}\label{app::algorithm}
The pseudo codes of training and generation algorithms are in Algorithms 1 and 2.

\begin{algorithm}[!h]
    \caption{Training algorithm of GraphVF}
    \label{alg:train}
    \begin{flushleft}
            \textbf{Input}: $\eta$ learning rate, $B$ batch size, $T$ maximum epoch number, Variational annealing hyperparameters $\beta_\mathrm{min}, \beta_\mathrm{max}$, use $\operatorname{Prod}(\cdot)$ as the product of elements across dimensions of a tensor
    \end{flushleft}
    \begin{flushleft}\textbf{Initial}: Parameters $\theta$ of GraphVF (Echnet, junction tree encoder, focal classifier, and Node/Edge/Position-MLP)\end{flushleft}
    \begin{algorithmic}[1] 
    \For{$t=1,..,T$} 
        \State $\beta = \beta_\mathrm{min} + (\beta_\mathrm{max} - \beta_\mathrm{min})\ \mathrm{sin^2}\left(\pi\frac{t}{T}\right)$ \Comment{$\beta$-annealing acc. to epoch number}
        \For{$b=1,...,B$}
            \State Sample a receptor-ligand pair from dataset, with receptor size $M$ and ligand size $N$
            \State Protein receptor $\mathcal{P} = (\tilde{V}, \tilde{E} )$, where $\tilde{V} = \left\{\left(\tilde a_i, \tilde r_i\right)\right\}_{i=1}^{M}$, and $\tilde{E} = \{\tilde b_{ij}\}_{i, j = 1}^{M}$
            \State Drug ligand $\mathcal{R} = ({V}, {E} )$, where ${V} = \left\{\left(a_i, r_i\right)\right\}_{i=1}^{N}$, and ${E} = \{b_{ij}\}_{i, j = 1}^{N}$
            \State Re-order $\mathcal R$ with ring-first graph traversal
            \State $(\vmu_\gR, \mSigma_\gR) = \text{JT-Encoder}(\mathcal R)$, where prior $Z\sim\mathcal N(\vmu_\gR, \mSigma_\gR)$
            \Comment{\textbf{2D Global}}
            \For{$i=1,...,N$} \Comment{\textbf{3D Autoregressive}}
                \State Construct sub-graph $\mathcal G_i := \tau(\mathcal P \cup \mathcal R_{1:i-1})$
                \If{i=1} 
                \State $f_i := $ Nearest receptor atom 
                \State $\boldsymbol{\hat{y}} = \text{one-hot}_M(f_i)$
                \State Predict focal score $\boldsymbol{y}$ for all receptor atoms $1:M$
                \Else 
                \State $f_i := i - 1$ 
                \State $\boldsymbol{\hat{y}} = \text{one-hot}_{i-1}(f_i)$
                \State Predict focal score $\boldsymbol{y}$ for all known ligand atoms $1:i-1$
                \EndIf
                \State $\mathbf{\tilde e}_{1:M}, \mathbf{e}_{1:i-1} = \mathrm{Echnet}(\mathcal G_i)$ \Comment{Encode 3D conformation}

                \State $\mathbf{x}_i^\mathrm{(node)} = a_i + \mathbf u,\ \mathbf u\sim \mathcal U[0,1)^{d^\mathrm{(node)}}$ \Comment{Atom type dequantization}
                \State $\mu_i^\mathrm{(node)}, \sigma_i^\mathrm{(node)} = \text{Node-MLP}(\mathbf{e}_{f_i})$
                \State $\mathbf{z}_i^\mathrm{(node)} = \left(\mathbf{x}_i^\mathrm{(node)} - \mu_i^\mathrm{(node)}\right) \odot \frac{1}{\sigma_i^\mathrm{(node)}}$

                \State $\mathbf{x}_{1:i-1, i}^\mathrm{(bond)} = b_{1:i-1, i} + \mathbf u,\ \mathbf u\sim \mathcal U[0,1)^{(i-1)\times {d^\mathrm{(bond)}}}$ \Comment{Bond type dequantization}
                \State $\mu_{1:i-1, i}^\mathrm{(bond)}, \sigma_{1:i-1, i}^\mathrm{(bond)} = \text{Bond-MLP}(\mathbf{e}_{1:i-1}, \mathbf{x}_i^\mathrm{(node)})$
                \State $\mathbf{z}_{1:i-1, i}^\mathrm{(bond)} = \left(\mathbf{x}_{1:i-1, i}^\mathrm{(bond)} - \mu_{1:i-1, i}^\mathrm{(bond)}\right) \odot \frac{1}{\sigma_{1:i-1, i}^\mathrm{(bond)}}$

                \State $\mathbf{x}_i^\mathrm{(pos)} = 
                \mathrm{Spherize}(\boldsymbol{r}_i; f_i, c_i, e_i)$
                \Comment{Spherize atom position to $f_i$}
                \State $\mu_i^\mathrm{(pos)}, \sigma_i^\mathrm{(pos)} = \text{Position-MLP}(\mathbf{e}_{f_i}, \mathbf{x}_i^\mathrm{(node)}, \mathbf{x}_{1:i-1, i}^\mathrm{(bond)})$
                \State $\mathbf{z}_i^\mathrm{(pos)} = \left(\mathbf{x}_i^\mathrm{(pos)} - \mu_i^\mathrm{(pos)}\right) \odot \frac{1}{\sigma_i^\mathrm{(pos)}}$
                
                \State $\mathcal{L}_i^\mathrm{(node)}  = -\log(\operatorname{Prod}(\mathcal N(\mathbf{z}_i^\mathrm{(node)}|\vmu_\gR, \mSigma_\gR))) -\log(\operatorname{Prod}(\frac{1}{\sigma_i^\mathrm{(node)}}))$
                \State $\mathcal{L}_{1:i-1, i}^\mathrm{(bond)}  = -\log(\operatorname{Prod}(\mathcal N(\mathbf{z}_{1:i-1, i}^\mathrm{(bond)}|\vmu_\gR, \mSigma_\gR))) -\log(\operatorname{Prod}(\frac{1}{\sigma_{1:i-1, i}^\mathrm{(bond)}}))$
                \State $\mathcal{L}_i^\mathrm{(pos)}  = -\log(\operatorname{Prod}(\mathcal N(\mathbf{z}_i^\mathrm{(pos)}|\vmu_\gR, \mSigma_\gR))) -\log(\operatorname{Prod}(\frac{1}{\sigma_i^\mathrm{(pos)}}))$
                \State $\mathcal{L}^{i, b}_\mathrm{flow} = \mathcal{L}_i^\mathrm{(node)} + \mathcal{L}_{1:i-1, i}^\mathrm{(bond)} + \mathcal{L}_i^\mathrm{(pos)}$ \Comment{Step-wise loss term for normalizing flow}
                \State $\mathcal{L}_\mathrm{focal}^{i,b} = \mathrm{BCELoss}(\boldsymbol{y}, \boldsymbol{\hat{y}})$ \Comment{Step-wise loss term for focal classifier}

            \EndFor

            \State $\mathcal{L}_\mathrm{KL}^b = D_\mathrm{KL}(\mathcal N(\vmu_\gR, \mSigma_\gR)||\mathcal N(\mathbf{0},\boldsymbol{I}))$ \Comment{Global loss term for variational distribution $Z$}
            \State $\mathcal{L}^{b}_\mathrm{total} = \frac{1}{N}\sum_{i=1}^{N}\left(\mathcal{L}^{i, b}_\mathrm{flow} + \mathcal{L}^{i, b}_\mathrm{focal}\right) + \beta \mathcal{L}_\mathrm{KL}^b$

        \EndFor
        \State $\theta \leftarrow \text{ADAM} (\sum_{b=1}^{B}\mathcal{L}^{b}_\mathrm{total}, \theta, \eta)$ \Comment{Parameter update}
    \EndFor
\end{algorithmic}
\end{algorithm}

\begin{algorithm}[!h]
    \caption{Generation algorithm of GraphVF}
    \label{alg:generate}
    \begin{flushleft}
         \textbf{Input}: $T$ number of protein receptors, $B$ number of drug ligands to generate for each receptor, $N$ maximum number of atoms in the generated ligand. Optional parameters $(\vmu_\rho, \mSigma_\rho)$ as the cue to certain desired property $\rho$, $(\mathbf{0},\boldsymbol{I})$ by default. 
   
    \end{flushleft}
    \begin{flushleft}\textbf{Initial}: Trained GraphVF model (Echnet, junction tree encoder, focal classifier, and Node/Edge/Position-MLP)\end{flushleft}
    \begin{algorithmic}[1] 
    \For{$t=1,..,T$} 
        \State Sample a protein receptor from dataset, with receptor size $M$
        \State Protein receptor $\mathcal{P} = (\tilde{V}, \tilde{E} )$, where $\tilde{V} = \left\{\left(\tilde a_i, \tilde r_i\right)\right\}_{i=1}^{M}$, and $\tilde{E} = \{\tilde b_{ij}\}_{i, j = 1}^{M}$
        \State $\mathrm{LigGen}_t\leftarrow [\ ]$
        \For{$b=1,...,B$}
            \State Drug ligand representation $\mathcal{R} := ({V}, {E} )$, initialized as empty
            \For{$i=1,...,N$} 
                \State Construct sub-graph $\mathcal G_i := \tau(\mathcal P \cup \mathcal R_{1:i-1}$)
                \State Predict focal score, sample focal atom $f_i$ from eligible atoms 
                \If{none eligible for $f_i$} \Comment{Signal for generation complete}
                    \State \textbf{break} inner loop
                \EndIf
                \State $\mathbf{\tilde e}_{1:M},\mathbf{e}_{1:i-1} = \mathrm{Echnet}(\mathcal G_i)$ \Comment{Encode 3D conformation}

                \State Sample $\left[\mathbf{z}_i^{\text {(node)}} ; \mathbf{z}_{1: i-1, i}^{\text {(bond)}} ; \mathbf{z}_i^{\text {(pos)}}\right] \sim \mathcal{N}\left(\boldsymbol{\mu}_{\mathcal{R}}, \boldsymbol{\Sigma}_{\mathcal{R}}\right)$ \Comment{Sample from latent space}
                \State $\mu_i^\mathrm{(node)}, \sigma_i^\mathrm{(node)} = \text{Node-MLP}(\mathbf{e}_{f_i})$ 
                \State $\mathbf{x}_i^\mathrm{(node)} = \sigma_i^\mathrm{(node)} \odot \mathbf{z}_i^\mathrm{(node)} + \mu_i^\mathrm{(node)}$ \Comment{Atom type generation}

                \State $\mu_{1:i-1, i}^\mathrm{(bond)}, \sigma_{1:i-1, i}^\mathrm{(bond)} = \text{Bond-MLP}(\mathbf{e}_{1:i-1}, \mathbf{x}_i^\mathrm{(node)})$
                \State $\mathbf{x}_{1:i-1, i}^\mathrm{(bond)} = \sigma_{1:i-1, i}^\mathrm{(bond)} \odot \mathbf{z}_{1:i-1, i}^\mathrm{(bond)} + \mu_{1:i-1, i}^\mathrm{(bond)}$ \Comment{Bond type generation}

                \State $\mu_i^\mathrm{(pos)}, \sigma_i^\mathrm{(pos)} = \text{Position-MLP}(\mathbf{e}_{f_i}, \mathbf{x}_i^\mathrm{(node)}, \mathbf{x}_{1:i-1, i}^\mathrm{(bond)})$
                \State $\mathbf{x}_i^\mathrm{(pos)} = \sigma_i^\mathrm{(pos)} \odot \mathbf{z}_i^\mathrm{(pos)} + \mu_i^\mathrm{(pos)}$ \Comment{Atom position generation}
                \State Derive $(a_i, \boldsymbol{r}_i)$ from $(\mathbf{x}_i^\mathrm{(node)}, \mathbf{x}_i^\mathrm{(pos)})$; $b_{1:i-1, i}$ from $\mathbf{x}_{1:i-1, i}^\mathrm{(bond)}$
                \State $V.\mathrm{append}(\{(a_i, r_i)\}); E.\mathrm{append}(\{b_{1:i-1, i}\})$ \Comment{Autoregressive ligand generation}
            \EndFor
            \State $\mathrm{LigGen}_t.\mathrm{append}(\mathcal{R})$
        \EndFor
    \EndFor
    \State \textbf{return} {$[\mathrm{LigGen}_1, \mathrm{LigGen}_2, ..., \mathrm{LigGen}_T]$}
\end{algorithmic}
\end{algorithm}

\section{Chemical structure illustration}
\label{app:chem-struc-illus}
\begin{figure}[!ht]
    \centering
    \includegraphics[width=0.7\linewidth]{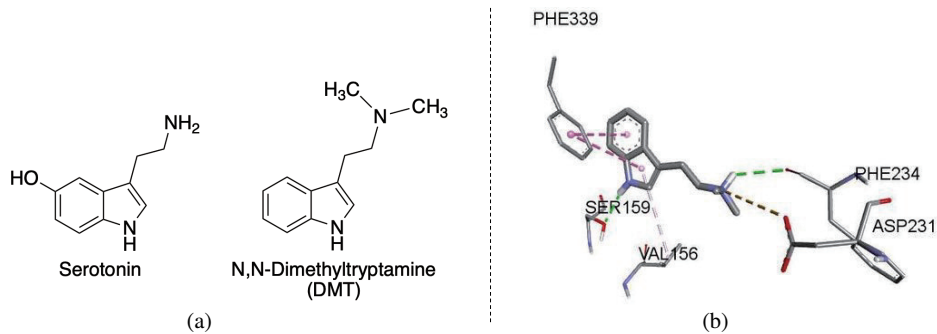}
    \caption{(a) Comparison between Serotonin and DMT structure; 
    (b) Binding pose of DMT with 5-$\mathrm{HT_{2A}}$, pay special attention to the interaction between $\mathrm{NHMe_2}$ and Asp-231.}
    \label{figure:pharmacophore}
    \vspace{-1.7em}
\end{figure}
\newpage

\section{Focal atom and the spherical coordinate system}
\label{app:focal-scs}
\subsection{Focal atom selection}
\label{app:focal-atom-selection}
During the \textbf{training/generation} process, focal scores are evaluated at each step. In the first step, there are no known ligand atoms, so the focal score is evaluated among the receptor molecules. In the following steps, focal atoms are only evaluated among the known ligand atoms. 

For the \textbf{generation} process, atoms become \textit{eligible} candidates of the focal atom, as long as their focal scores exceed a given threshold (set to 0.5 in our implementation). The focal atom of that step is then randomly sampled from these eligible candidates. In the case when no atoms are eligible for the focal atom, the generation process is called to an end. 

For the \textbf{training} process, we use a teacher-forcing strategy. That is to say, we rely on an expert trajectory (
{curated from the ring-first traversal algorithm, described in Appendix~\ref{app:ring-first-traversal}}) to select our focal atom. In the first step, the focal atom is set to be the receptor atom that is nearest to the drug ligand. In the $i$-th step ($i>1$), the focal atom is set to be the $(i-1)$-th atom in the expert trajectory. The ground truth label for focal score is set to either 1 (focal) or 0 (non-focal), and we use a mean-reduced BCELoss to evaluate the discrepancy between the predicted focal score and the ground-truth label. 

\subsection{Construction of the spherical coordinate system (SCS)}
\label{app:construction-scs}

We construct a spherical coordinate system (SCS) around the focal atom $f$ and its nearest 2 neighboring atoms $c$ and $e$. Formally, given the Cartesian coordinates of reference atoms $(\boldsymbol{r}_f, \boldsymbol{r}_c, \boldsymbol{r}_e)$, we want to express the position of an arbitrary atom $i$ in spherical coordinates $(d_i, \theta_i, \phi_i)$. Suppose $\boldsymbol{r}_i$ is the Cartesian coordinates of atom $i$, $\boldsymbol{n}_1$ is the normal vector of plane $(\boldsymbol{r}_f, \boldsymbol{r}_c, \boldsymbol{r}_i)$, and  $\boldsymbol{n}_2$ is the normal vector of plane $(\boldsymbol{r}_f, \boldsymbol{r}_c, \boldsymbol{r}_e)$, 
then 
\begin{equation}
\begin{cases}
d_i = \left\|\boldsymbol{r}_i-\boldsymbol{r}_f\right\|_2,  & d_i \ge 0 \\
\theta_i = \left \langle \boldsymbol{r}_i - \boldsymbol{r}_f, \boldsymbol{r}_c - \boldsymbol{r}_f \right \rangle,  & \theta_i \in[0, \pi] \\ 
\phi_i = \left \langle \boldsymbol{n}_1, \boldsymbol{n}_2 \right \rangle,  & \phi_i \in[-\pi, \pi]
\end{cases}
\end{equation}
Conversely, we can also render the Cartesian coordinates of $i$ from its spherical coordinates: 
\begin{equation}
    (x_i, y_i, z_i) = \boldsymbol{r}_i = \boldsymbol{r}_f+\frac{d_i (\boldsymbol{r}_c-\boldsymbol{r}_f)\cos \theta_i}{\left\|\boldsymbol{r}_c-\boldsymbol{r}_f\right\|_2^2}+\frac{d_i (\boldsymbol{r}_{e, {{\phi}_i}}-\boldsymbol{r}_{e, c f})\sin \theta_i}{\left\|\boldsymbol{r}_{e, {{\phi}_i}}-\boldsymbol{r}_{e, c f}\right\|_2^2}, 
\end{equation}
where $\boldsymbol{r}_{e, c f}$ is the coordinate of the projection of $e$ on the line $(\boldsymbol{r}_f, \boldsymbol{r}_c)$, and $\boldsymbol{r}_{e, \phi_i}$ is the coordinate of $e$ after rotating the plane $(\boldsymbol{r}_f, \boldsymbol{r}_c, \boldsymbol{r}_e)$ along the line $(\boldsymbol{r}_f, \boldsymbol{r}_c)$ by the torsion angle $\phi_i$. We define this operation as $h: (d_i, \theta_i, \phi_i) \mapsto (x_i, y_i, z_i)$. Note that this transformation between the Cartesian and spherical coordinates is SE(3)-equivariant. Namely, for any orthogonal matrix $Q \in \mathbb{R}^{3 \times 3}$ and translation vector $\boldsymbol{b} \in \mathbb{R}^3$: 
\begin{align}
& h\left(d_i, \theta_i, \phi_i ; Q \boldsymbol{r}_f+\boldsymbol{b}, Q \boldsymbol{r}_c+\boldsymbol{b}, Q \boldsymbol{r}_e+\boldsymbol{b}\right) \label{eq:Cartesian-LHS}\\
= & Q \boldsymbol{r}_f+\boldsymbol{b}+\frac{d_i \cos \theta_i\left(Q \boldsymbol{r}_c-Q \boldsymbol{r}_f\right)}{\left\|Q \boldsymbol{r}_c-Q \boldsymbol{r}_f\right\|_2^2}+\frac{d_i \sin \theta_i\left(Q \boldsymbol{r}_{e, \phi_i}-Q \boldsymbol{r}_{e, c f}\right)}{\left\|Q \boldsymbol{r}_{e, \phi_i}-Q \boldsymbol{r}_{e, c f}\right\|_2^2} \\
= & Q \boldsymbol{r}_f+\boldsymbol{b}+Q \frac{d_i \cos \theta_i\left(\boldsymbol{r}_c-\boldsymbol{r}_f\right)}{\left\|\boldsymbol{r}_c-\boldsymbol{r}_f\right\|_2^2}+Q \frac{d_i \sin \theta_i\left(\boldsymbol{r}_{e, \phi_i}-\boldsymbol{r}_{e, c f}\right)}{\left\|\boldsymbol{r}_{e, \phi_i}-\boldsymbol{r}_{e, c f}\right\|_2^2} \\
= & Q\left[\boldsymbol{r}_f+\frac{d_i \cos \theta_i\left(\boldsymbol{r}_c-\boldsymbol{r}_f\right)}{\left\|\boldsymbol{r}_c-\boldsymbol{r}_f\right\|_2^2}+\frac{d_i \sin \theta_i\left(\boldsymbol{r}_{e, \phi_i}-\boldsymbol{r}_{e, c f}\right)}{\left\|\boldsymbol{r}_{e, \phi_i}-\boldsymbol{r}_{e, c f}\right\|_2^2}\right]+\boldsymbol{b} \\
= & Q h\left(d_i, \theta_i, \phi_i ; \boldsymbol{r}_f, \boldsymbol{r}_c, \boldsymbol{r}_e\right)+\boldsymbol{b} .\label{eq:Cartesian-RHS}
\end{align}

\subsection{Equivariance of the generation process}
\label{app:equivariance}

The intuition behind preserving equivariance during the generation process is: when the rest part of a molecule moves in space, the newly generated part of that molecule should move accordingly. Formally, we have the following theorem: 

\textbf{Theorem 1 (SE(3)-equivariant generation).} At the $i$-th ($i = 1, ..., N$) generation step, the generation probability is equivariant to any orthogonal matrix $Q \in \mathbb{R}^{3 \times 3}$ and translation vector $\boldsymbol{b} \in \mathbb{R}^3$: 
\begin{equation}
    p\left(Q \boldsymbol{r}_i+\boldsymbol{b} \mid A_i, B_i, R_i Q^T+\boldsymbol{1b}^T\right)=p\left(\boldsymbol{r}_i \mid A_i, B_i, R_i\right),
    \label{eq:theorem-equivariance-gen}
\end{equation}
where $A_i$ is the types of all known atoms, $B_i$ is the types of all known bonds, and $R_i$ is the Cartesian coordinates of all known atoms. 

$\boldsymbol{1^\circ}$ To prove the above theorem, we first show that the spherical coordinates $(d_i, \theta_i, \phi_i)$ are SE(3)-invariant to flow transformation. By combining Equations \ref{eq:encode-subgraph} and \ref{eq:flow_forward}, we can derive the following clean-cut equation: 
\begin{equation}
    (d_i, \theta_i, \phi_i) = g(\mathbf{z}^a_i, \mathbf{z}^b_{1:i-1, i}, z^d_i, z^\theta_i, z^\phi_i; A_i, B_i, R_i), 
\end{equation}
where $g$ is an invertible flow transformation, parameterized by $A_i, B_i, R_i$ through EchNet. Since EchNet perceives 3D geometry only through relative distances, $g$ is thus SE(3)-invariant: 
\begin{equation}
    g(\mathbf{z}^a_i, \mathbf{z}^b_{1:i-1, i}, z^d_i, z^\theta_i, z^\phi_i; A_i, B_i, R_i Q^T+\boldsymbol{1b}^T) = g(\mathbf{z}^a_i, \mathbf{z}^b_{1:i-1, i}, z^d_i, z^\theta_i, z^\phi_i; A_i, B_i, R_i). \label{eq:scs-invariant}
\end{equation}

$\boldsymbol{2^\circ}$ We further show that Cartesian coordinates are SE(3)-equivariant to flow transformation under the same set of reference atoms. For Equation~\ref{eq:scs-invariant}, we substitute its LHS into Equation~\ref{eq:Cartesian-LHS}, and its RHS into Equation~\ref{eq:Cartesian-RHS}: 
\begin{align}
    & h\left(g(\mathbf{z}^a_i, \mathbf{z}^b_{1:i-1, i}, z^d_i, z^\theta_i, z^\phi_i; A_i, B_i, R_i Q^T+\boldsymbol{1b}^T) ; Q \boldsymbol{r}_f+\boldsymbol{b}, Q \boldsymbol{r}_c+\boldsymbol{b}, Q \boldsymbol{r}_e+\boldsymbol{b}\right) \\
    = & Q h\left(g(\mathbf{z}^a_i, \mathbf{z}^b_{1:i-1, i}, z^d_i, z^\theta_i, z^\phi_i; A_i, B_i, R_i) ; \boldsymbol{r}_f, \boldsymbol{r}_c, \boldsymbol{r}_e\right)+\boldsymbol{b} .
\end{align}
We define a short-hand composite function $g^r := h\circ g$, and the resultant equation unequivocally shows that Cartesian coordinates are SE(3)-equivariant to $g^r$ under the same $f, c, e$: 
\begin{equation}
    g^r(\mathbf{z}^a_i, \mathbf{z}^b_{1:i-1, i}, z^d_i, z^\theta_i, z^\phi_i; A_i, B_i, R_i Q^T+\boldsymbol{1b}^T) = Qg^r(\mathbf{z}^a_i, \mathbf{z}^b_{1:i-1, i}, z^d_i, z^\theta_i, z^\phi_i; A_i, B_i, R_i) + \boldsymbol{b} = Q\boldsymbol{r}_i + \boldsymbol{b}. \label{eq:Cartesian-Equivariant}
\end{equation}

$\boldsymbol{3^\circ}$ Finally, since both sides of Equation~\ref{eq:Cartesian-Equivariant} share the same underlying distribution $[\mathbf{z}^a_i; \mathbf{z}^b_{1:i-1, i}; z^d_i; z^\theta_i; z^\phi_i]\sim \mathcal N(\vmu_\gR, \mSigma_\gR)$, we eventually come to the SE(3)-equivariance of our generation process, formulated as Equation~\ref{eq:theorem-equivariance-gen}. \textbf{Theorem 1} is thus proved. 

\section{Ring-first traversal algorithm}
\label{app:ring-first-traversal}
\begin{figure}[!ht]
    \centering
    \includegraphics[width=0.8\linewidth]{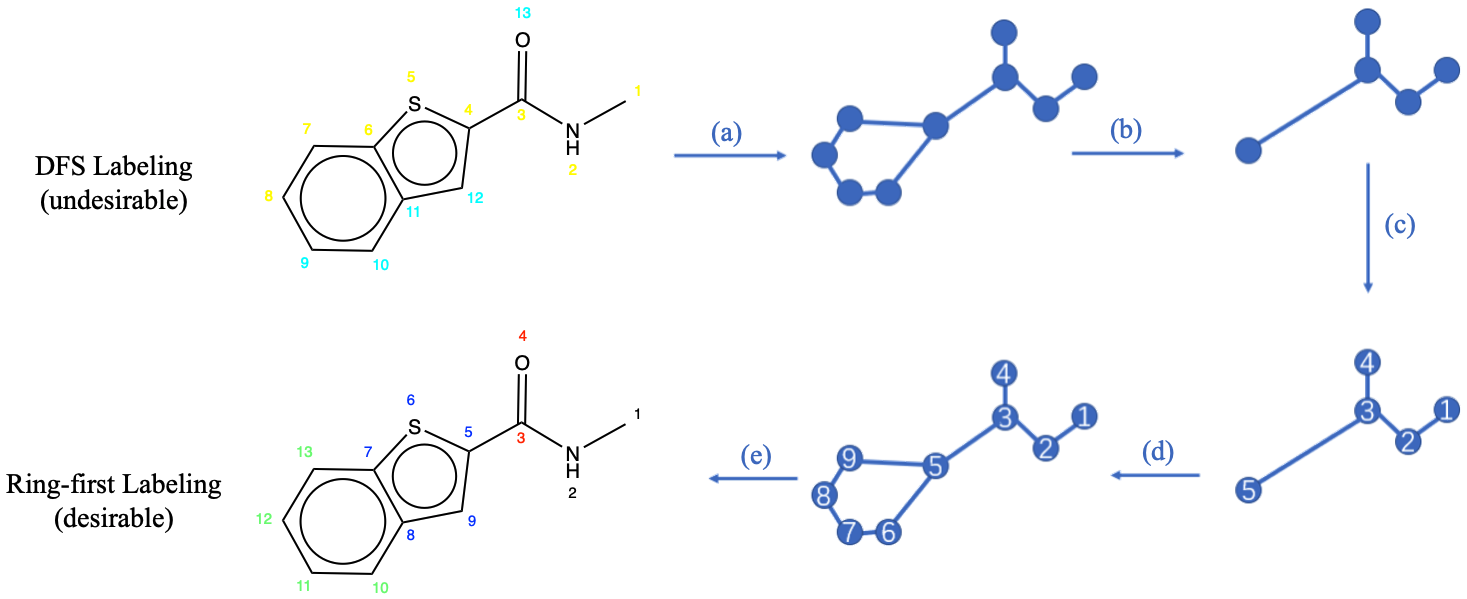}
    \caption{Overview of the ring-first traversal algorithm, from step (a) to (e). The desirable ring-first labeling is demonstrated in the down-left, and the undesirable DFS labeling is in the up-left. Consecutive labels are marked with the same color. }
    \label{figure:ring-first-illus}
\end{figure}
For our auto-regressive model, we curate an expert trajectory for training (Appendix~\ref{app:focal-atom-selection}), using the ring-first traversal algorithm. Chemical structures are typically composed of rings and functional groups, and the ring-first traversal algorithm tries to label atoms from the same ring/functional group consecutively to preserve the local semantics of the chemical structure \textbf{during training}. For example, in the left side of Figure~\ref{figure:ring-first-illus}, the ring-first labeling is a desirable delineation of the different rings/functional groups existing in the molecule (benzene, thiophene, and carbonyl), but the DFS labeling mixes them up. 

We propose a minimal-ring Floyd algorithm to implement the ring-first strategy. As shown in Figure~\ref{figure:ring-first-illus}, it consists of 5 steps: 
\begin{enumerate}[label=(\alph*)]
    \item Find the minimal ring via Floyd's algorthm~\citep{poj1734, weisstein2008floyd}, and substitute it into a single point; 
    \item Repeat step (a), until there is no ring in the graph, \textit{i.e.} the graph has become a tree; 
    \item Apply depth-first search (DFS)~\citep{tarjan1972depth} to label the tree nodes; 
    \item If a tree node has once been rings, expand it back into a ring. Label the nodes in the same ring consecutively. Offset the labeling of successive nodes accordingly; 
    \item Repeat step (d), until the original molecular graph structure is restored. The resultant labeling is the desirable ring-first labeling. 
\end{enumerate}

This ring-first traversal algorithm has a time complexity of $\mathcal{O}(n^5)$, where $n$ is the number of nodes in the molecular graph (excluding hydrogen atoms). To accelerate the algorithm, we refactor our code with Numba~\citep{lam2015numba} to allow just-in-time compilation and achieve a marked $\sim 50\times $ acceleration. We have thus been able to improve our training procedure using ring-first traversal, without losing efficiency. The detailed implementation of the ring-first traversal algorithm can be found in our code repository.

\end{document}